\documentclass{aa}  
\usepackage{lscape}
\usepackage{graphicx}
\usepackage{threeparttable}
\usepackage{txfonts}
\begin{document}

   \title{A homogeneous analysis of globular clusters from the APOGEE survey with the BACCHUS code. I. The Northern clusters}
\titlerunning{BACCHUS and the 10 GCs}

   \author{T. Masseron \inst{1,2}
          \and
         D.A.Garc\'ia-Hern\'andez \inst{1,2}
          \and 
         Sz. M\'esz\'aros\inst{3,4}
          \and 
          O. Zamora\inst{1,2}         
         \and 
          F. Dell’Agli\inst{1,2}
          \and
          C. Allende Prieto\inst{1,2}
           \and
          B. Edvardsson \inst{5}
          \and 
          M. Shetrone\inst{6}
         \and
          B. Plez\inst{7}
          \and
          J. G. Fernández-Trincado\inst{8,9,10}
          \and
          K. Cunha \inst{11,12}
          \and
          H, Jönsson \inst{13}
          \and
          D. Geisler \inst{9,14,15}
          \and
          T.~C.~Beers \inst{16}
          \and 
          R.~E.~Cohen \inst{17}
          }
          
   \institute{Instituto de Astrof\'isica de Canarias, E-38205 La Laguna, Tenerife, Spain
   	\and 
Departamento de Astrof\'isica, Universidad de La Laguna, E-38206 La Laguna, Tenerife, Spain\\
              \email{tmasseron@iac.es}
         \and
ELTE E\"otv\"os Lor\'and University, Gothard Astrophysical Observatory, Szombathely, Hungary
        \and
    Premium Postdoctoral Fellow of the Hungarian Academy of Sciences
             \and 
Theoretical Astrophysics, Department of Physics and Astronomy, Uppsala University, Box 516, SE-751 20 Uppsala, Sweden
             \and 
McDonald Observatory, University of Texas at Austin, Fort Davis, TX 79734, USA 
 \and 
Laboratoire Univers et Particules de Montpellier, Universit\'e de Montpellier, CNRS, 34095, Montpellier Cedex 05, France
  \and 
  Instituto de Astronom\'ia y Ciencias Planetarias, Universidad de Atacama, Copayapu 485, Copiap\'o, Chile
  \and
Departamento de Astronom\'ia, Casilla 160-C, Universidad de Concepci\'on, Concepci\'on, Chile
\and
Institut Utinam, CNRS UMR 6213, Universit\'e Bourgogne-Franche-Comt\'e, OSU THETA Franche-Comt\'e, Observatoire de Besan\c{c}on,  BP 1615, 25010 Besan\c{c}on Cedex, France
\and
Observatório Nacional/MCTI, Rua Gen. José Cristino, 77, 20921-400 Rio de Janeiro, Brazil
\and
Steward Observatory, University of Arizona Tucson 85719, USA
\and
Lund Observatory, Department of Astronomy and Theoretical Physics, Lund University, Box 43, SE-22100 Lund, Sweden
 \and
 Instituto de Investigaci\'on Multidisciplinario en Ciencia y Tecnolog\'ia, Universidad de La Serena. Avenida Ra\'ul Bitrán S/N, La Serena, Chile
\and
Departamento de F\'isica y Astronom\'ia, Facultad de Ciencias, Universidad de La Serena. Av. Juan Cisternas 1200, La Serena, Chile
 \and
Department of Physics and JINA Center for the Evolution of the Elements, University of Notre Dame, Notre Dame, IN 46556, USA      
\and
Space Telescope Science Institute, 3700 San Martin Drive, Baltimore, MD 21210, USA
             }

   \date{Received ; accepted }

 
  \abstract
  {}   
   {We aim at providing abundances of a large set of light and neutron-capture elements homogeneously analyzed and covering a wide range of metallicity to constrain globular cluster (GC) formation and evolution models.}
   {We analyze a large sample of 885 GCs giants from the APOGEE survey. We used the Cannon results to separate the red giant branch and the asymptotic giant branch stars, not only allowing for a refinement of surface gravity from isochrones, but also providing an independent H-band spectroscopic method to distinguish stellar evolutionary status in clusters. We then use the BACCHUS code to derive metallicity, microturbulence, macroturbulence and many light-element abundances as well as the neutron-capture elements Nd and Ce for the first time from the APOGEE GCs data. }
   {Our independent analysis helped us to diagnose issues regarding the standard analysis of the APOGEE DR14 for low-metallicity GC stars.  Furthermore, while we confirm most of the known correlations and anti-correlation trends (Na-O, Mg-Al, C-N), we discover that some stars within our most metal-poor clusters show an extreme Mg depletion and some Si enhancement but at the same time show some relative Al depletion, displaying a turnover in the Mg-Al diagram. These stars suggest that Al has been partially depleted in their progenitors by very hot proton-capture nucleosynthetic processes. Furthermore, we attempted to quantitatively correlate the spread of Al abundances with the global properties of GCs. We find an anti-correlation of the Al spread against clusters metallicity and luminosity, but the data do not allow to find clear evidence of a dependence of N against metallicity in the more metal-poor clusters.}
   {Large and homogeneously analyzed samples from on-going spectroscopic surveys unveil unseen chemical details for many clusters, including a turnover in the Mg-Al anti-correlation, thus yielding new constrains for GCs formation/evolution models.}


   \maketitle
%
%
 
\section{Introduction}
The existence of multiple populations in Globular Clusters (GC) can be unambiguously observed in appropriate colour-magnitude diagrams \citep[e.g.][and reference therein]{Milone2017} and the variations in colours are associated to abundances variations \citep[see e.g.][]{Monelli2013,Meszaros2018}. The colour indices sensitive to multiple populations have such a sensitivity because their band-pass includes the spectral features that are changing, for example $\rm C_{UBI}$ or the Milone et al. "magic trio" with WFC3 UVIS.  Some broadband colours (V-I in most cases) are mostly insensitive to such (C,N,O etc.) variations, but most colours indices are sensitive to metallicity in various ways.
Indeed, it is known for a long time that some elemental abundances vary from star to star within the clusters. The most observed elements showing abundance variations are C, N, O, Mg, Na and Al \citep[see past reviews and references therein]{Smith1987,Kraft1994,Gratton2004,Gratton2012}. Si has been more recently revealed to vary in some clusters \citep[e.g.][]{Yong2005,Carretta200917clusters}. Ca is another element that has been revealed to show spread in some clusters \citep{Marino2009}. Finally, K has been observed to vary in only two clusters: NGC~2808 and NGC~2419 \citep{CohenKirby2012,Mucciarelli2015}. Regarding neutron capture elements, a few clusters show significant dispersion \citep[ including M~22, M~15, M~92 and M4]{Marino2009,Sobeck2011,Roederer2011,Shingles2014}. Last but not least, some colour indices may suggest some He enhancement \citep{Milone2015}. Correlation or anti-correlation between those elements have been observed and provide hints to decipher their origins. While there is a consensus on a hot H-burning nucleosynthesis source, a broad range of polluters have been proposed to explain those chemical trends: fast  rotating   massive   stars,  massive asymptotic giant branch (AGB) stars, intermediate-mass binaries and supermassive stars \citep[see discussion of the various models by][]{Renzini2015,Charbonnel2016}. \\
Nevertheless, all those models try to establish a universal scenario for the formation of GCs. One way to distinguish those scenarios may come from confronting the various abundance trends and spread against global properties of the clusters, as already observationally attempted by \citet{Carretta200917clusters} and \citet{Milone2017} and theoretically predicted by \citet{DellAgli2018} and \citet{Szecsi2018}. To do this, large samples of homogeneously analyzed stars are required to allow to compare trends or (anti-)correlations against cluster properties and thus draw an overall picture of GC formation. But, as emphasized by \citet{BastianLardo2018}, to  date,  there  have  only  been  a  few  stars  in  a  handful  of  GCs  that  have  been  fully characterized in terms of their chemistry.

In fact, the largest published spectroscopic homogeneous analysis of GCs stars to date is from \citet{Carretta2009NaO} who gathered nearly 1400 stellar spectra from the VLT/GIRAFFE spectrograph, but they study only Na and O because of relatively limited resolution. In parallel, \citet{Carretta200917clusters} measured several elements over 17 GCs, but the sample is only of 202 stars. Given the potential scientific impact of such studies, large spectroscopic surveys have now dedicated generous amount of telescope time for the observation of GCs. \citet{Pancino2017} makes use of the Gaia-ESO survey \citep{Gilmore2012} and studied 510 stars over 9 clusters, but limited their conclusions to Mg and Al elements. \citet{Meszaros2015} presented an independent analysis of 428 Northern cluster stars using spectra published in the 10th data release of the SDSS~III-Apache Point Observatory Galactic Evolution Experiment (APOGEE) \citep{Ahn2014}. However, this analysis suffered from larger than expected uncertainties in the fundamental C and N abundances. Nowadays, the fourteenth data release (DR14) of the SDSS~IV/APOGEE2 survey \citep{Gunn2006,Zasowski2017,Majewski2017,Blanton2017} contains one of the largest samples of GCs stars with high enough resolution and signal-to-noise ratio to allow the determination of abundances for many key elemental abundances for GCs studies. However, the extreme elemental abundances of GC stars are still uncertain in DR14 standard analysis as demonstrated by \citet{Jonsson2018}. 

Given the larger than expected errors in the derivation of C and N abundances by \citet{Meszaros2015} and the APOGEE Stellar Parameters and Chemical Abundances Pipeline (ASPCAP) \citep{Holtzman2018}, we independently revisit here the analysis of APOGEE spectra from 10 Northern GCs and nearly double the sample compared to \citet{Meszaros2015}'s work (now 885) as more stars have been observed by APOGEE since then. The current analysis includes measurement of C, N, O, Mg, Al, Si, K and Ca as well as the neutron-capture elements Ce and Nd for the first time in GCs since those elements have been shown to be measurable in the APOGEE spectra \citep{Hasselquist2016,Cunha2017}.
 \begin{figure*}
	\includegraphics[angle=-90,width=0.75\linewidth]{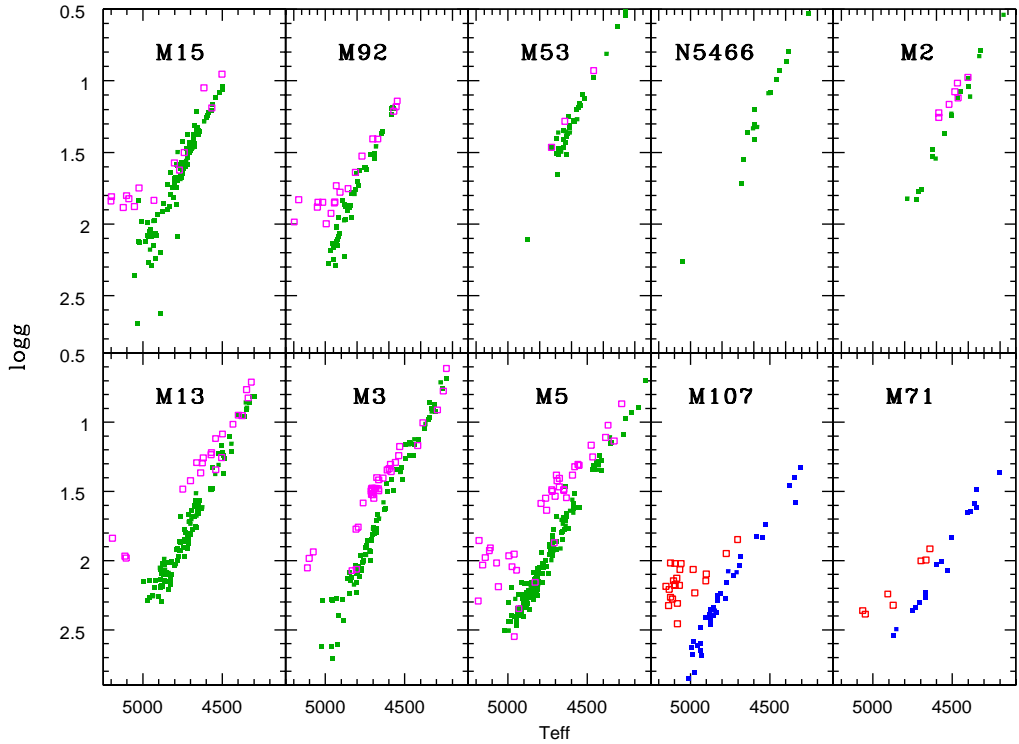}
   \centering
    \caption{Effective temperatures and surface gravities derived by the Cannon (DR14) for the sample stars. Open magenta squares are RHB/eAGB stars and green full squares are RGB stars as determined from photometry except for M~107 and M~71, for which the separation has been done from spectroscopic parameters only. The figure illustrates the potentiality of separating RGB and RHB/eAGB solely from the Cannon results.}
              \label{fig:Teffvslogg}%
    \end{figure*}
  \begin{figure*}[h!]
	\includegraphics[angle=-90,width=0.75\linewidth]{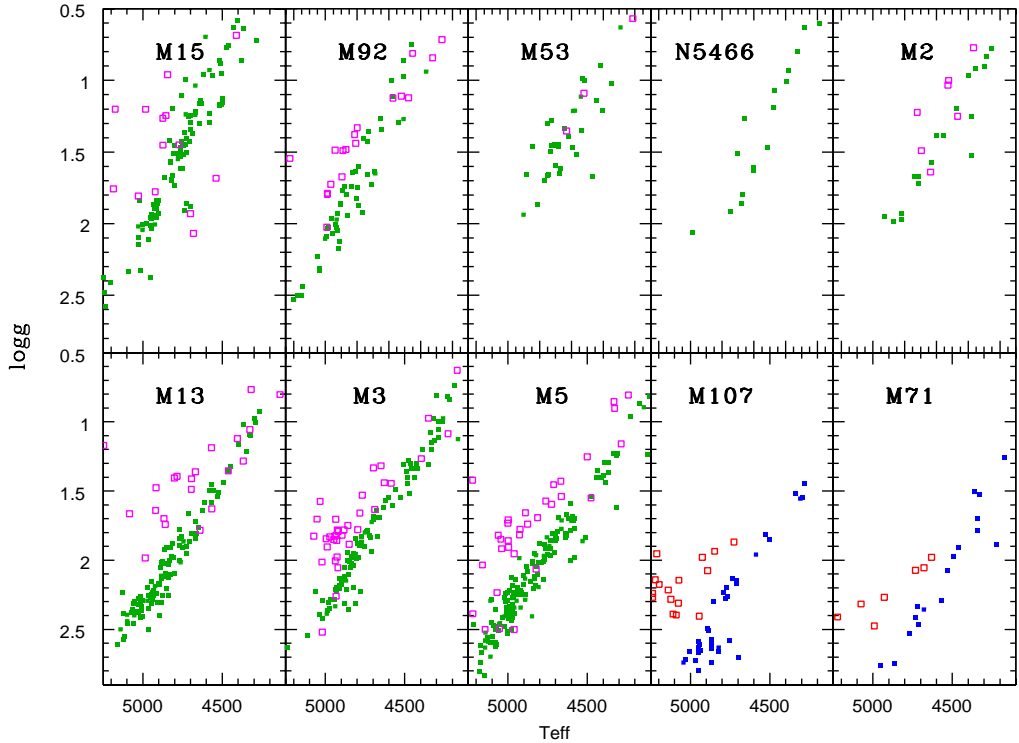}
   \centering
    \caption{ Effective temperatures and surface gravities derived by ASPCAP (DR14) for the sample stars. Open magenta squares are RHB/eAGB stars and green full squares are RGB stars as determined from photometry except for M~107 and M~71, for which the separation has been done from spectroscopic parameters only. In comparison to Fig.~\ref{fig:Teffvslogg}, this figures illustrates that the Cannon has more precise results when it comes to separating the RGB/RHB/eAGB, but as shown in \citet{Jonsson2018}, the ASPCAP parameters are more accurate when comparing to independently analyzed local disk stars.}
              \label{fig:TeffvsloggASPCAP}%
    \end{figure*}

   \begin{figure}
	\includegraphics[angle=-90,width=0.95\columnwidth]{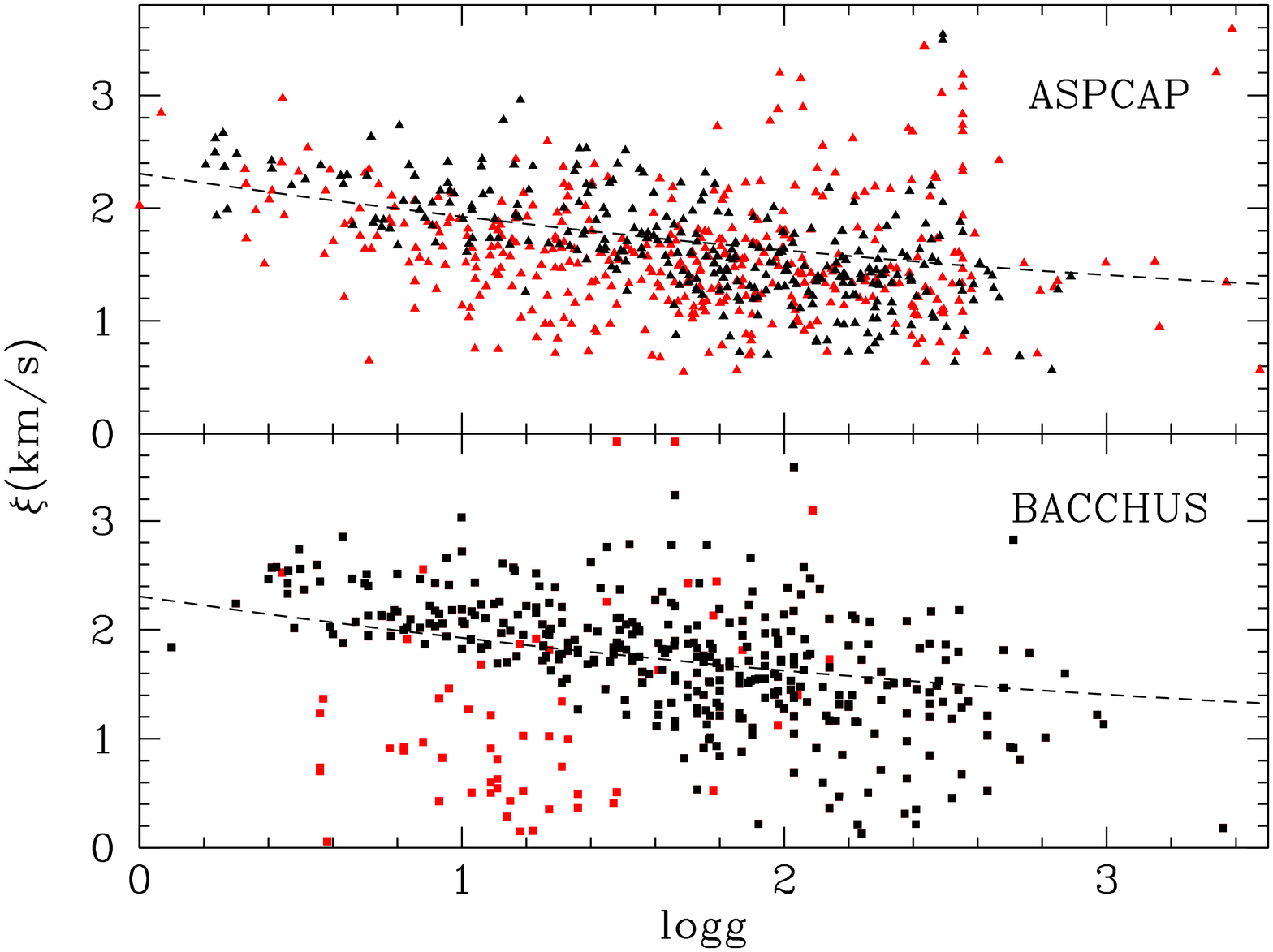}
   \centering
   \caption{Microturbulence velocity values obtained for the sample stars by the ASPCAP pipeline (upper panel) and by our preliminary run with the BACCHUS code (lower panel). Red points indicate stars with [Fe/H] $<$ -1.5 and black points stars with [Fe/H] $>$ -1.5. The dashed line show the relation derived in the optical by the Gaia-ESO survey assuming T$\rm_{eff}$= 4500~K and  [Fe/H]=-1.5. Both works show a good correspondence with the optical data for the metal-rich stars, but a larger dispersion is obtained for low-metallicity stars.}
              \label{fig:microturb}%
    \end{figure}
    
\section{Data}
\subsection{Sample selection}
The general steps of target selection were the same as used by \citet{Meszaros2015}. After selecting radial velocity members, we cut stars that were beyond the tidal radius given by \citet{Harris2010}, and stars that had an offset larger than $\pm$0.4 dex from the average ASPCAP DR14 metallicity. Gaia DR2 data have not been used for the member selection because distances are currently not accurate enough for most of these clusters. The new member selection provided a factor of two more stars, 885 compared to 428, than \citet{Meszaros2015}, most of these new stars being observed after DR10 was published. 

The corresponding spectra used in the present work were obtained from the SDSS~IV/APOGEE2 survey DR14 \citep{Abolfathi2018}. The spectra have been reduced and combined following \citet{Holtzman2015,Holtzman2018}, but the spectral normalization has been done within our own code.

\section{Analysis}
We use the Brussels Automatic Code for Characterizing High accUracy Spectra (BACCHUS) \citep{Masseron2016} to determine metallicity, microturbulence and macroturbulence/$v\sin i$ as well as abundances (C, N, O, Mg, Al, Na, Ca, Si, K, Nd and Ce). The code runs on the fly the spectral synthesis code Turbospectrum \citep{Plez2012} and rely on a very large grid of MARCS model atmospheres \citep{Gustafsson2008}, extended over various C values (-1.0$<$[C/Fe]$<$+1.0) in addition to more standard parameters (T$\rm _{eff}, \log$g, metallicity). Note that although the grid also contains a range of [$\alpha$/Fe] ($\alpha$ being O, Ne, Mg, Si, S, Ar, Ca, and Ti), it does not contain models with variation in O, Mg and Si, but fixed Ne, Ar, Ca and Ti. Therefore we decided to limit the grid such that [$\alpha$/Fe]=+0.4. We also stress that the atmospheric models we use do not take into account the variations in Na or Al, nor possibly in He. The atomic linelist used is from DR14 \citep[see][for a description of the linelist building]{Shetrone2015} complemented with Nd and Ce from \citep{Hasselquist2016} and \citet{Cunha2017}. The molecular linelists include recent updates from \citet{Sneden2014} (CN),  \citet{Li2015,HITRAN2016} (CO) and \citet{Brooke2016} (OH).

The sequence of the spectral analysis procedure is the following: after having fixed the effective temperatures and surface gravities (see Sec.\ref{sec:Tefflogg}), the code determines the macroturbulence based on a selection of Si~I lines (simultaneously to adjust the Si abundance). Then, metallicity is determined from Fe~I lines with macro- and microturbulence parameters fixed. Because of the importance of molecular lines in the H band, we successively determine the O, C and N abundances that dominate the molecular line strengths. Finally, we derived the other element abundances. The overall process was iterated twice to ensure self-consistency and ran over a couple of days for the whole sample on our HTCondor sytem.

\subsection{Effective temperature and surface gravity}\label{sec:Tefflogg}
\citet{Jonsson2018} illustrated that the ASPCAP DR14 analysis leads to an excellent precision in abundances for disk stars and first generation GC stars, lending support for the associated values of effective temperature and gravity. Nevertheless, \citet{Jonsson2018} has also demonstrated that AS<PCAP DR14 analysis do not provide satisfying results for second generation GC stars. Moreover, as detailed in Sec.~\ref{sec:abundances}, we found that the temperatures provided by ASPCAP for GC stars introduce a bias in the oxygen abundances. Our BACCHUS analysis does not determine T$\rm_{eff}$ and $\log$g, so we derive those parameters similarly to \citet{Meszaros2015}, i.e. temperatures are inferred from photometry and surface gravities from isochrones. We computed the effective temperatures by using the J-Ks relation from \citet{Gonzalez2009}, while the reddening values were selected from the Harris catalog \citep{Harris2010} and kept as constant for each cluster.

In order to assign correctly the surface gravities with isochrones, a distinction between early asymptotic giant branch (eAGB) or red horizontal branch (RHB),  and red giant branch (RGB) stars must be made. This has been done by using colour-colour diagrams based on the most accurate ground-based photometry currently available using the same method as \citet{GarciaHernandez2015}. However, such photometric data are only available for the most metal-poor clusters, and not for the two most metal-rich clusters in our sample, M~71 and M~107. For these two clusters, we chose to use the Cannon \citep{Ness2015} analysis of DR14 parameters. 

While the accuracy of this machine learning algorithm is limited to that of its training set, it has the advantage of exploiting all of the information that may be present in the stellar spectra, thus improving the precision of the stellar parameters determination. We show in Fig.~\ref{fig:Teffvslogg} that the parameters derived by the Cannon allow us to disentangle efficiently the RGB from the RHB/eAGB stars. In comparison, we show in Fig.~\ref{fig:TeffvsloggASPCAP} a similar diagram but with the ASPCAP DR14 calibrated parameters. Although the ASPCAP results also allow to separate the two giant branches, the Cannon results show it more clearly. Consequently, we have separated empirically the RHB/eAGB and the RGB stars in M~107 and M~71 using the Cannon results.  This provides a new method for separating solely from H-band spectroscopy the evolutionary status of giants in clusters, or in mono-age populations in general. In contrast, we found that the effective temperatures and the surface gravities provided by the Cannon are much less accurate than the ASPCAP values, thus the latter are preferred for parameter determinations of globular clusters. We note that a few stars in M~5, M~13, M~3 and M~15 have been apparently  misclassified by the Cannon and/or ASPCAP. However, it is not clear at this stage of this study whether it is due to an error in the Cannon determination of the parameters, or due to an error in the photometric classification. More work would be needed to clarify the situation of those few outliers and it is beyond the scope of this paper.

Once the evolutionary status is assigned, the surface gravities are determined by using the set of isochrones from the Padova group \citep{bertelli01,Marigo2017}. We use 12 billion years isochrones for all GCs, except for  M~107 and M~71 for which we use 10 billion years. Regarding metallicities, we adapt the isochrones to the metallicities of the \citet{Harris2010} catalog.

\subsection{Micro- and macroturbulence}
   \begin{figure}
	\includegraphics[angle=-90,width=\columnwidth]{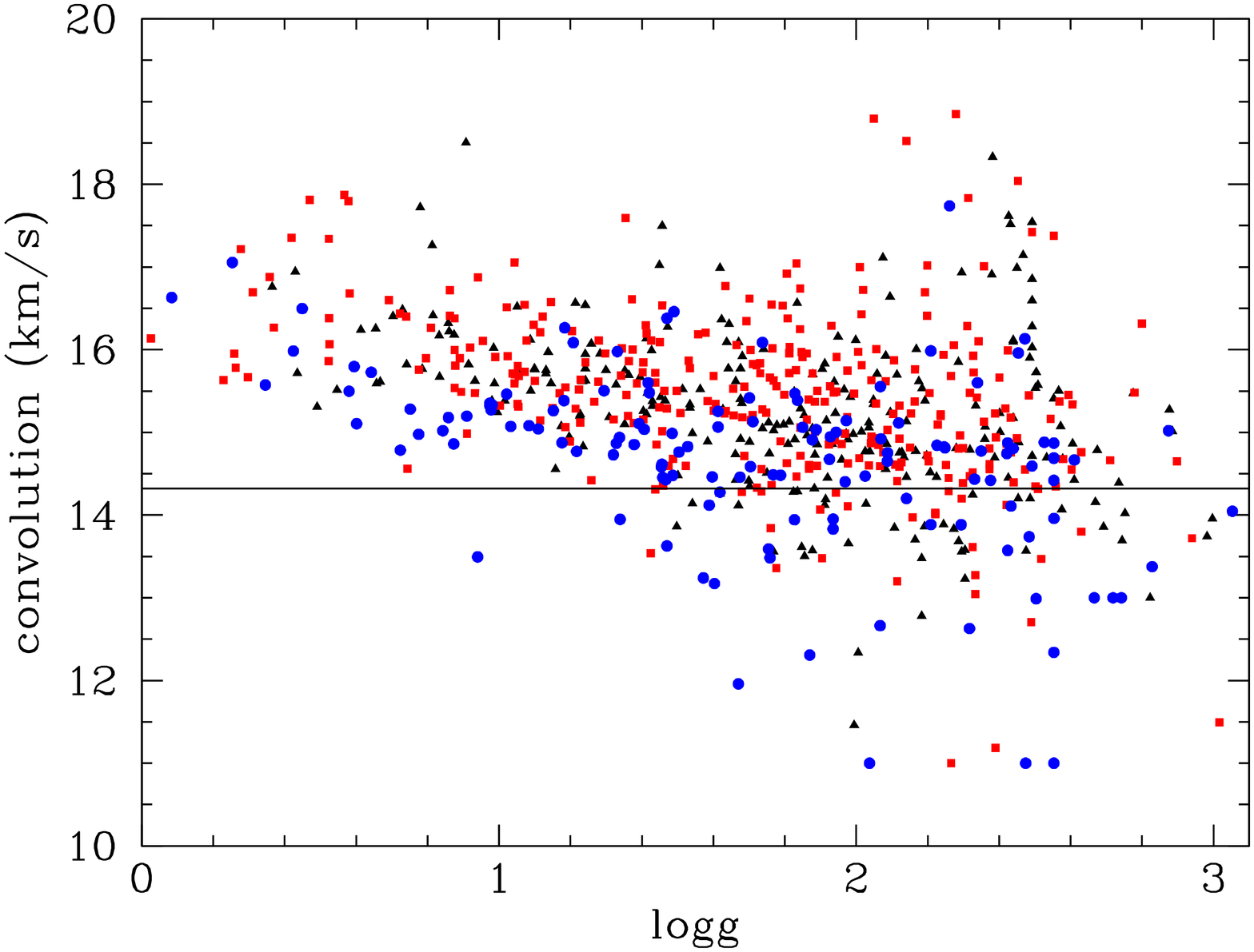}
	\includegraphics[angle=-90,width=\columnwidth]{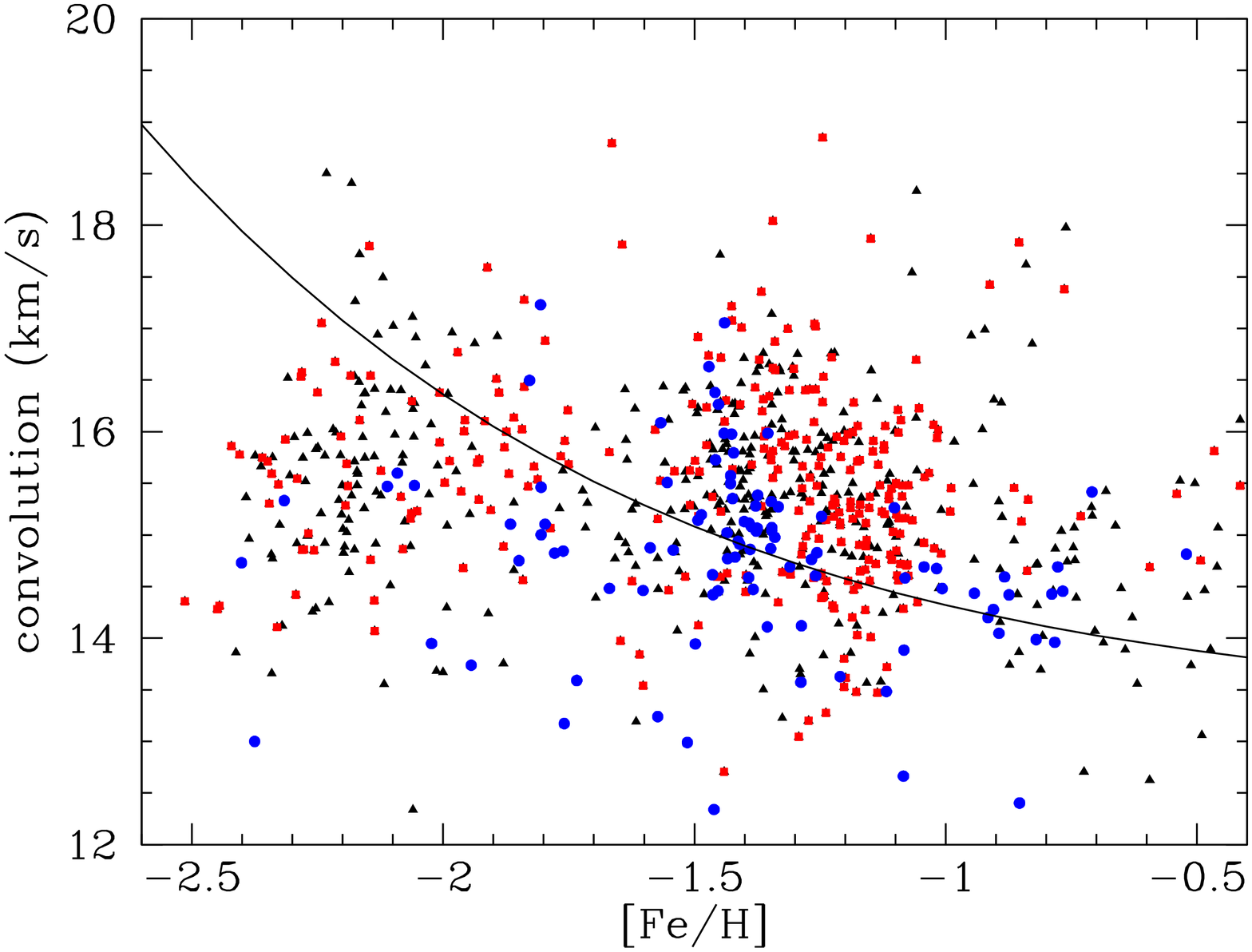}
   \centering
   \caption{Convolution values obtained for the sample stars against surface gravities (upper panel) and metallicity (lower panel). Red squares are spectra with a low mean fiber number ($<$100), blue points are spectra with a high mean fiber number ($>$200), and black triangles are the others. The continuous line correspond to the macroturbulence relation adopted in APOGEE DR14 quadratically summed with an average resolution of 13.3 km/s (R$\sim$22,500). }
              \label{fig:convol}%
    \end{figure}

Microturbulence velocities have been fixed to the relation obtained in the optical by the Gaia-ESO survey \citep[such as described by ][]{Smiljanic2014}. However, in order to check the validity of this relation in the H-band infrared spectra of APOGEE, we let the microturbulence parameter free in a preliminary run. The code determines the microturbulence by cancelling the trend of abundances against equivalent widths of a set of Fe~I lines. Fig.~\ref{fig:microturb} shows the results obtained with BACCHUS and compares them to the values derived in APOGEE DR14 and to the relation of the Gaia-ESO survey. All the relations agree well and show in particular an expected dependence with surface gravity. Nevertheless, the dispersion of the microturbulence values increase significantly for the most metal-poor stars ([Fe/H]< -1.5). This is due to the fact that the lines used for the microturbulence relation become very weak and very sensitive to random uncertainties such as noise or continuum placement, and thus increasing the uncertainties of the microturbulence relation. Because our study contains a large number of very low-metallicity stars, we decided to fix the microturbulence velocity to the optical relation from Gaia-ESO in order to reduce the impact of the growing error at low metallicity on the abundances.

Concerning the macroturbulence value, the BACCHUS code does not derive it directly, but rather derives the value of the spectral convolution necessary to match the observed line profile after the thermal, radiative, collisional and microturbulence broadening have been taken into account in the synthesis. This convolution parameter approximately represents the quadratic sum of instrumental resolution, projected velocity of the star ($v\sin i$) and macroturbulence. To determine it, the BACCHUS code iterates over a series of selected lines until it finds convergence between abundances measured from the few points around the core of the line (being sensitive to the convolution value) and from the equivalent width of the lines (insensitive to the convolution). For this particular analysis, we use a Gaussian kernel for the convolution and probe it over a set of clean Si~I lines (those lines being strong enough in all our metal-poor stars in contrast to Fe~I lines, see next Subsection).

In Fig.~\ref{fig:convol}, we show the resulting convolution values derived from our procedure against $\log g$ and metallicity. We highlighted the average fiber number of each target in the figure. It is remarkable that we obtained higher convolution values on average for stars with a larger fiber number than the ones with a lower fiber number. This is fully consistent  with the fact that the APOGEE spectrograph has a variable resolution depending on the fiber \citep{Nidever2015}. Although the BACCHUS code is making some approximations regarding the macroturbulence and instrument line profile broadening, the recovery of the fiber impact on the spectra demonstrates the ability of the code to account for the various source of line broadening parameters when deriving the stars abundances.

In the same Figure, we compare our convolution values to the adopted value for macroturbulence from DR14 APOGEE data release \citep{Holtzman2018}. \citet{Holtzman2018} found a dependence only in metallicity in DR14. Our results tend to rather show a dependence in $\log g$, consistently with previous findings from \citet{Hekker2007,Gray2008}, but we do not find a dependence on metallicity. We remind the reader that the macroturbulence velocity relation derived in \citet{Holtzman2018} is based on the whole APOGEE sample, which is dominated by more metal-rich stars than present in our GC sample and that may be responsible for some of the disagreement. In any case, we stress that  if the macroturbulence is really overestimated at low metallicity, we expect that all abundances and parameters derived from the ASPCAP pipeline to be significantly affected for all stars with metallicities below -2.0.

\subsection{Metallicity}
\begin{figure}
	\includegraphics[angle=-90,width=\columnwidth]{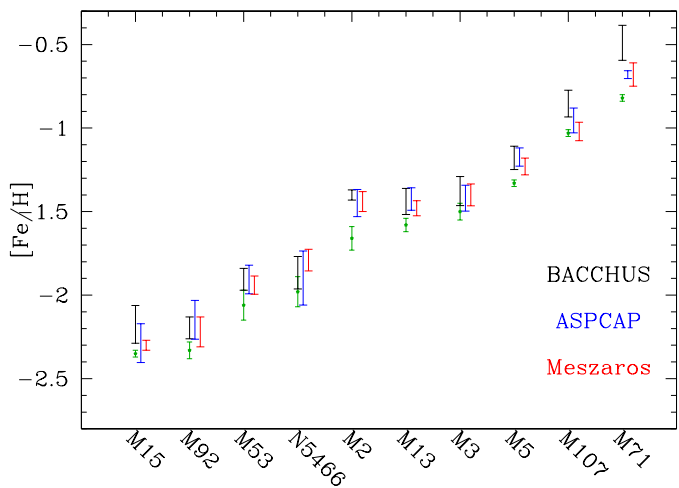}
   \centering
   \caption{Median metallicities obtained by the present work, APOGEE DR14 and \citet{Meszaros2015} for the 10 same GCs stars. Errorbars indicate the star-by-star metallicity rms. Green points are literature values \citep[][]{Carretta2009Fespread}. }
              \label{fig:metallicity}%
    \end{figure}

Once the macroturbulence, microturbulence, T$\rm _{eff}$ and  $\rm \log$g have been fixed, the metallicity is derived from a selection of three Fe~I lines. Fig.~\ref{fig:metallicity} shows the median metallicity obtained for each cluster and compared to various literature values. The agreement  between all the studies is good overall. While there is evidence for metallicity variations within some peculiar clusters \citep[e.g.  $\omega$Cen, NGC~6934 ][]{Norris1995,Marino2018}, the relatively low star-to-star metallicity dispersion as shown in Fig.~\ref{fig:metallicity} is consistent with the conclusions of \citet{Carretta2009Fespread}, claiming that most of the clusters can be considered mono-metallic regarding the Fe abundance. 
However, we find a systematic offset of $\sim$0.1dex in metallicity compared to the optical values for all clusters, and so do other studies of the APOGEE spectra \citet{Meszaros2015,Holtzman2018}.  \citet{Mucciarelli2015NLTE} already demonstrated that NLTE effects in optical Fe~I lines in  M~2 stars affect the metallicity. In the H-band we also strongly suspect 3D-NLTE effects, notably because 1D-LTE synthesis of a very high resolution of Arcturus spectrum does not provide a satisfying fit for Fe~I line profiles. A more extended discussion will be given in a forthcoming paper. It is for this reason that we carefully select only three Fe~I lines which are as insensitive as possible to NLTE effects to determine the metallicities.

In any case, such an offset in metallicity does not affect our discussion, because we mostly focus on the relative trends or ranges of abundance-abundance diagrams, cluster by cluster.

   \begin{figure}
	\includegraphics[angle=-90,width=\columnwidth]{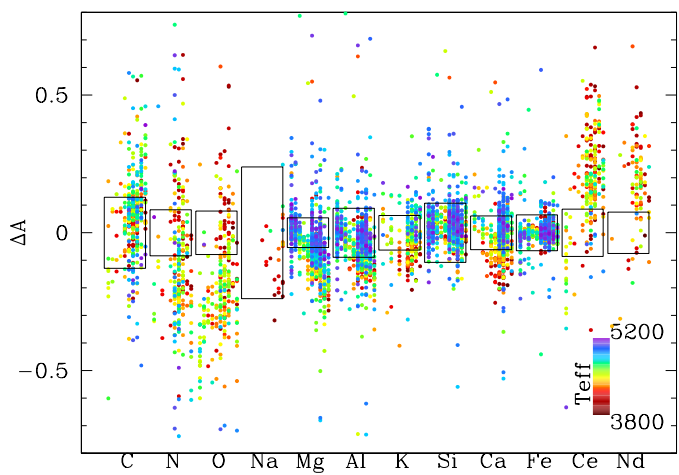}
    \centering
   \caption{Difference in abundances obtained by BACCHUS using APOGEE T$\rm_{eff}$ and $\rm \log$g with those obtained using our parameters. The points are colour-coded by temperature and the points are ordered by cluster metallicity for each element. The squares represent the mean random line-by-line scatter uncertainties of each element. Systematic uncertainties are comparable to random uncertainties except for C, N, O, Ce and Nd.}
              \label{fig:errors}%
    \end{figure}
%
\begin{figure}
	\includegraphics[width=\columnwidth]{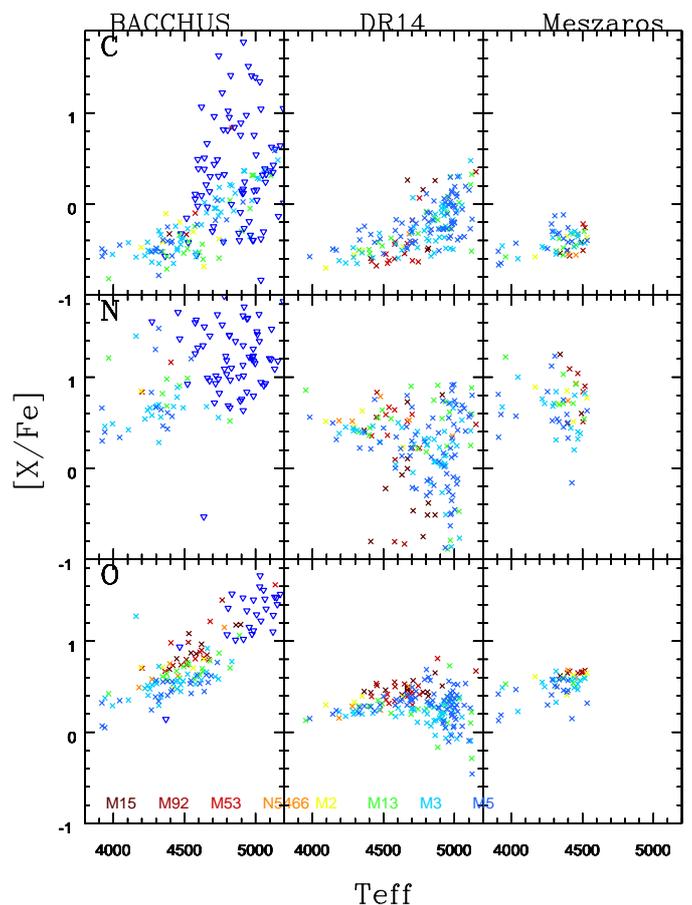}
    \centering
   \caption{The C, N and O abundances as a function of effective temperature for the same stars as a function of effective temperature for three studies \citep[this work, first column, uncalibrated APOGEE DR14, second column, and][third column]{Meszaros2015}. Blue triangles indicates upper limits. Our results show that C, N and O abundances have mostly upper limits for $\rm T_{eff} >$ 4600~K, whereas the APOGEE DR14 results do not provide upper limits. \citet{Meszaros2015} chose not to provide C, N, O values for $\rm T_{eff} >$ 4500~K. There is a significant trend in our O determinations compared to the two other studies for which we do not have a clear explanation (see text).}
              \label{fig:CNOvsTeff}%
    \end{figure}
    
   \begin{figure}
	\includegraphics[width=\columnwidth]{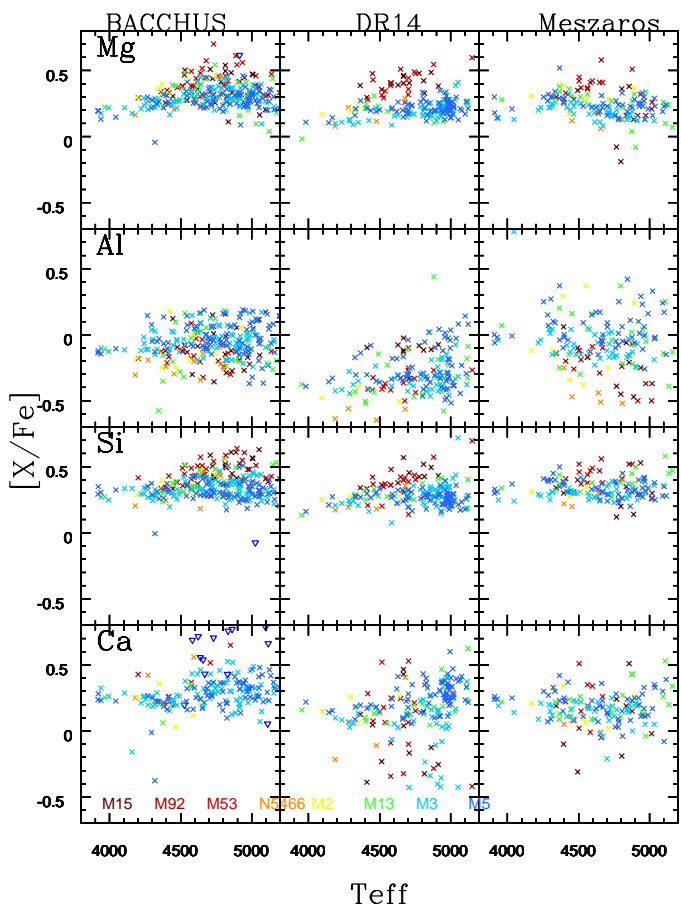}
	    \centering
   \caption{Comparison Mg, Al, Si and Ca abundances for the same stars as a function of effective temperature for three studies \citep[this work, first column, uncalibrated APOGEE DR14, second column, and][, third column]{Meszaros2015}. Blue  triangles are upper limits. To clear up the diagrams with intrinsic clusters abundances variations, only stars of the first population ([Al/Fe] < 0.2) are shown. For clarity sake, the results for two more metal-rich clusters M~71 and M~107 have been disregarded in this figure.}
              \label{fig:MgAlSiCavsTeff}%
    \end{figure}
   \begin{figure}
	\includegraphics[width=\columnwidth]{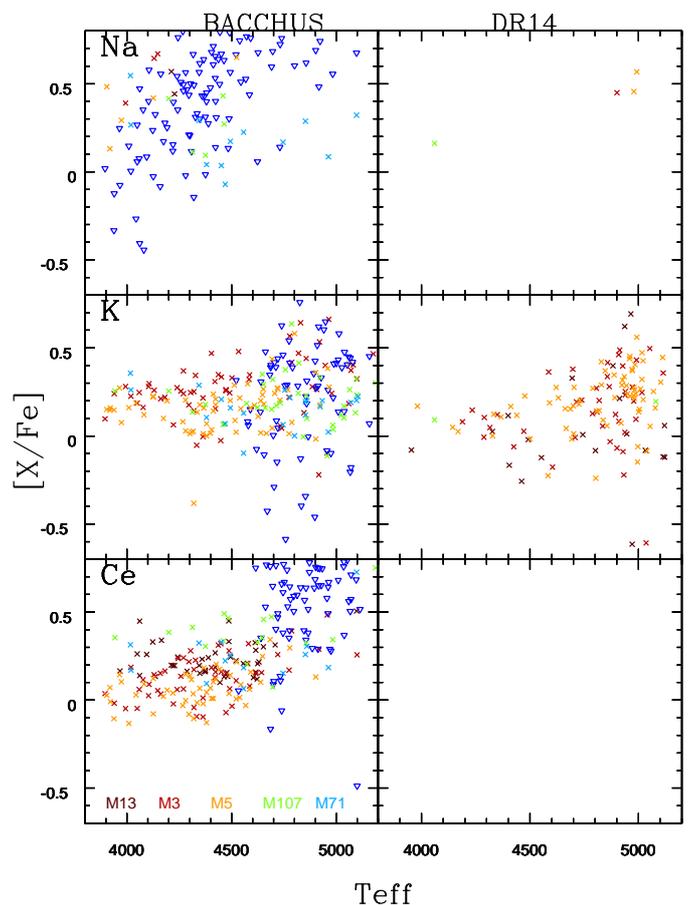}
    \centering
   \caption{Comparison Na, K and Ce abundances as a function of effective temperature for two studies (this work, first column, uncalibrated APOGEE DR14, second column). Blue  triangles are upper limits. The more metal-poor clusters are not shown because those clusters contains mostly upper limits values.}
              \label{fig:NaKCevsTeff}%
    \end{figure}

   \begin{figure}
	\includegraphics[width=0.95\columnwidth]{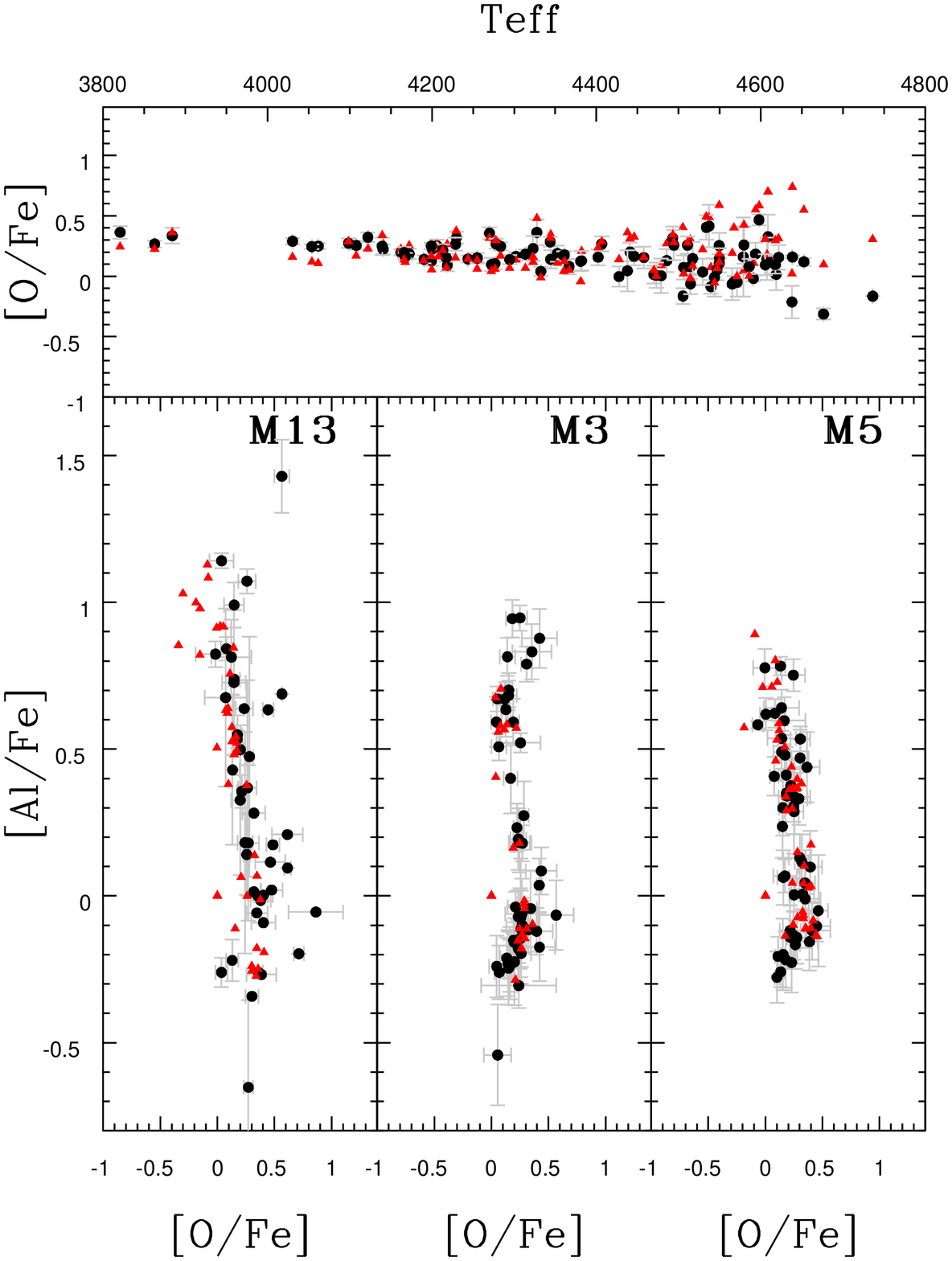}
    \centering
   \caption{O abundances as a function of effective temperature (upper panel) and against Al abundances for three of our clusters (bottom panels) determined by the BACCHUS code with the APOGEE DR14 uncalibrated T$\rm _{eff}$ and $\log$g (black dots) and by the ASPCAP pipeline (uncalibrated values, red triangles). The oxygen trend is slightly positive, at the opposite of Fig.~\ref{fig:CNOvsTeff}. Moreover, almost no O dispersion nor anti-correlation is observed, which cast some doubts on the stellar parameters.}
              \label{fig:AlFevsOFe_3clusters_ASPCAP}%
    \end{figure}

\subsection{Abundances}\label{sec:abundances}
Abundances have been derived by the code on a line-by-line basis. The lines used for  abundance determination are displayed in Tab.~\ref{tab:lines}. Note that the BACCHUS code automatically adjusts the window/mask on a star-by-star and line-by-line basis \citep[see the manual for more details][]{Masseron2016}. Therefore we provide only central wavelengths in Table~\ref{tab:lines}.  Among the several abundance indicators that the code offers (equivalent width, core line intensity or $\chi^2$) and their respective flags, we selected the abundances corresponding to the minimum $\chi^2$, but still use the other methods to reject any suspicious line. The final abundance is the mean of the abundances of the non-flagged lines. 

\begin{table}
\begin{tabular}{ll}
\hline \\
element &  wavelength ($\rm\AA_{air}$) \\
\hline \\ 
C (CO)  & 15578.0 15775.5 15978.7 16185.5 16397.2 16614.0 \\
        & 16836.0 17063.0 17448.6 17456.0 \\
N (CN)  &  15119.0 15210.2 15222.0 15228.8 15242.5 15251.8 \\
        &  15309.0 15317.6 15363.5 15410.5 15447.0 15462.4 \\
        & 15466.2 15495.0 15514.0 15581.0 15636.5 15659.0 \\
        & 15825.7 15391.0 15569.0 15778.5 16052.9 16055.5 \\
        & 16650.0 16704.8 16714.5 16872.0 16909.4 \\
O (OH)  &  15391.0 15569.0 15778.5 16052.9 16055.5 16650.0 \\
        & 16704.8 16714.5 16872.0 16909.4 \\
Na~I    & 16373.9 16388.8 \\
Mg~I    &  15740.7 15749.0 15765.7 \\
Al~I    &  16719.0 16750.6 \\
Si~I    &  15361.2 15376.8 15557.8 15827.2 15833.6 15884.5 \\
        & 15960.1 16060.0 16094.8 16129.0 16163.7 16170.2 \\
        &  16215.7 16241.8 16680.8 16828.2 \\
K~I     &  15163.0 15168.3 \\
Ca~I    &  16136.8 16150.8 16155.2 16157.4 16197.1 \\
Fe~I    & 15207.5 15294.6  15662.0   \\
Ce~II   &  15277.6 15784.7 15829.8 15958.4 15977.1 16327.3 \\
        &  16376.4 16595.1 16522.5 \\
Nd~II   &  15284.4    15368.1    15912.2    15977.9 16053.6    16262.0 \\               &  16303.7    16382.9    16558.2    16634.6  \\
\hline \\
\end{tabular}
\caption{Lines used in this work for abundance determination.}
\label{tab:lines}
\end{table}
\subsubsection{Errors}
Systematic errors have been evaluated by comparing the abundances obtained using the uncalibrated effective temperature and surface gravities from APOGEE DR14 and running the BACCHUS code following the same procedure as described above. The corresponding results are reported for each star of the sample in Fig.~\ref{fig:errors} and compared to the mean random uncertainties (the latter being derived from the line-by-line abundance dispersion). In this figure, systematic uncertainties are comparable to random uncertainties for most of the elements. However, they appear significantly larger for C, N, O, Nd and Ce. This is due to the high sensitivity of their abundances to effective temperature and/or surface gravity. Indeed, the APOGEE DR14 parameters tend to show systematic parameter differences with optical spectroscopy, as highlighted by \citet{Jonsson2018}, affecting the abundance determination of the most sensitive elements. Nevertheless, in Fig.~\ref{fig:errors} we illustrate the imàct of systematic errors on the abundances of Na, Mg, K, Si, Ca and Fe as well as on C, N, and O, by using another set of effective temperature or surface gravity. These effects are relatively small compared to the intrinsic variations observed in GCs (see Sec.~\ref{sec:Discussion}), and thus do not affect our conclusions regarding those elements.

\subsubsection{Upper limits}
Determining upper limits of abundances can be particularly useful when lines become too weak to be accurately measured. Actually, many lines used for abundance determination  become particularly weak at the metallicity of many GC spectra. The BACCHUS code is able to determine for each line an upper limit to abundances by comparing the variance of the observed spectrum to the behaviour of the line strength in the synthetic spectra. The upper limit corresponds to the abundance where the line intensity is comparable to the variance, and is consequently in BACCHUS sensitive to stellar parameters, as well as spectral resolution or  signal-to-noise ratio. An illustration of the importance of flagging and determining upper limits is given in Fig.~\ref{fig:CNOvsTeff}. In this figure, we can observe that most of the upper limits for C, N and O elements are set for stars with effective temperatures above 4600~K (100~K higher than what \citet{Meszaros2015} have used). This naturally occurs because the strength of the molecular lines used for those elements quickly weakens as temperature increases. However, while a flagging system is clearly missing in the ASPCAP pipeline \citep{GarciaPerez2016}, and it is unclear to which extent the pipeline can actually measure those elements. Furthermore, to avoid any bias in the interpretation and discussion of our results, we will not show C, N and O abundances for stars with $\rm T_{eff} >$ 4600~K.

\subsubsection{Comparison with literature}
Figures \ref{fig:CNOvsTeff}, \ref{fig:MgAlSiCavsTeff} and \ref{fig:NaKCevsTeff}  show the resulting abundances as a function of effective temperature for three distinct analysis of the same APOGEE spectra: this work, the uncalibrated ASPCAP DR14 results, and those from \citet{Meszaros2015}. For clarity, we do not show all the sample in these diagrams, but rather select the most relevant stars or clusters to highlight any residual trends or offsets. \\
Apart from the stars with upper limits, the abundances we derive agree fairly well with previous studies. Nevertheless, in Fig.~\ref{fig:CNOvsTeff} we notice that the N values derived by the ASPCAP pipeline for the DR14 data do not reach values higher than 1.0. This is issue is due to the limits of the model grid,  that will be enlarged for the next 16th data release of SDSS. Nevertheless, we note that such extreme values in N and C are only expected in specific cases such as GCs stars \citep[e.g. ][]{ Meszaros2015} or peculiar N-rich stars  \citep{FernandezTrincado2016,Schiavon2017,Trincado2017,Trincado2018}. \citet{Meszaros2015} have extended the limit of [N/Fe] up to +1.5, but still restricting [C/Fe] $\geq$ -0.75, which made the abundances of the most C-poor and N-rich stars unreliable.

Furthermore, none of the measured elements seem to show a temperature dependence consistent with the results in literature, except for C, N and O. While the C trend is probably related to expected changes in the evolution of giants, N should rather increases with decreasing temperatures while O is not expected to remain constant with temperature. Moreover, both the \citet{Meszaros2015} and APOGEE DR14 results show a slight trend in [O/Fe] as a function of T$\rm_{eff}$, but our is significantly stronger. Several hypothesis for such a trend can be invoked: 
\begin{itemize}
\item i) our chosen T$\rm_{eff}$-color transformation may be poorly calibrated for metal-poor GCs stars \citep{Gonzalez2009}. Given the extreme sensitivity of O in the H-band as demonstrated by  \citet{Jonsson2018}, the O trend may be a residual from this colour-T$\rm_{eff}$ approximation. However, our temperature scale is similar to \citet{Meszaros2015}, thus should lead to a similar trend which, is not the case 
\item ii) we are using spherical stellar atmospheres, whereas APOGEE DR14 and \citet{Meszaros2015} used plane parallel models. The differences basically lead to cooler layer in the outer layers in the spherical case compared to the plane parallel, where molecular lines form. Consequently, the OH molecule from which O is measured is stronger in the synthesis with spherical models and thus lead to lower O abundances. While this may explain the difference between our study and previous analysis, the fact that we are using an improved approximation for radiative transfer and model atmosphere still does not explain why we obtain a trend. 
\item iii) it has been demonstrated that 3D corrections affect the OH lines in the H-band at very low-metallicity \citep{Dobrovolskas2015}, such that O abundances are overestimated  in a 1D analysis. But this effect goes towards the opposite direction to what is observed, where the coolest giants with larger 3D effects should have even lower O than what is shown in Fig.~\ref{fig:CNOvsTeff} . 
\item iv) last, we use model atmosphere with $\alpha$ elements composition (including O, Mg, and Si), as well as Na and Al,  fixed. But in globular clusters, O, Si and Mg are known to vary independently. Moreover, Al and Na are also known to vary in GCs. Those elements are non-negligible electron donors, which could also affect the atmospheric structure. We have tested the impact of changing the initial $\alpha$ elements content by 0.5 dex, but no significant change has been obtain regarding the trend of O versus effective temperature. 
\end{itemize}
Finally, we have derived O abundances with the BACCHUS code adopting the calibrated temperatures and surface gravities in APOGEE DR14. In Fig.~\ref{fig:AlFevsOFe_3clusters_ASPCAP}) we show these abundances as a function of the APOGEE uncalibrated temperatures and against Al. With the temperatures provided by APOGEE DR14, we can reproduced very well the APOGEE DR14 results and we do not observe anymore any significant trend with T$\rm_{eff}$ as we do with our photometric temperatures. But those measurements do not exhibit any oxygen variations in clusters and the well-established Al-O anti-correlation. This is clearly in contradiction with all findings from previous GCs cluster studies. Knowing that OH is very temperature sensitive in the H band, we interpret such paradox by the fact that the ASPCAP is biasing the effective temperatures in order to obtain O abundances such that the built-in relation [O/$\alpha$]=0 is satisfied. This interpretation is very well in line with the comparison of ASPCAP results against literature by \citet{Jonsson2018}. Therefore, although there may still be some systematic uncertainties leading to a trend in O against temperature, we conclude that using our photometric temperatures provides more realistic results than using the ASPCAP DR14 ones for GCs study. 

While an increasing trend of N with decreasing temperature is expected due to internal CN cycling along the RGB, our data rather seem to show the opposite (middle panel of Fig.~\ref{fig:CNOvsTeff}). \citet{Masseron2015} demonstrate that CN lines (thus N abundances) are extremely sensitive to the O abundances. Therefore, it is likely that the trend obtained is related to the O trend.

In any case, whenever this trend in O is due to a measurement bias or not, there is also still the possibility that this trend is real. Indeed, \citet{Dantona2007} and \citet{DiCriscienzo2018} predicted that O depletion may occur along the RGB when stars are enriched in He as this could be the case in GC stars \citep{Lardo2018,Milone2018}. Therefore, without a clear understanding of the reason behind the O trend, we chose not to correct it. Moreover, as shown in Sec.~\ref{sec:Discussion}, the observed intrinsic spreads in O and in N are larger than their amplitudes. Therefore, although this may affect the precision of our abundances it does not affect our conclusions. 

%
   \begin{figure}[h!]
	\includegraphics[angle=-90,width=0.9\columnwidth]{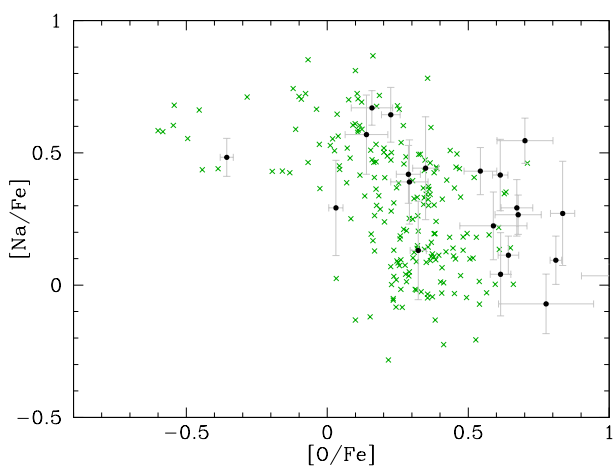}
   \centering
   \caption{Na abundances as a function of O (black points). Although there are only few stars in our sample where Na could be measured, the Na and O abundances follow the expected anti-correlation as measured by \citet{Carretta2009NaO} (green crosses) in the optical. }
              \label{fig:NavsO}%
    \end{figure}
   \begin{figure*}
	\includegraphics[angle=-90,width=0.7\linewidth]{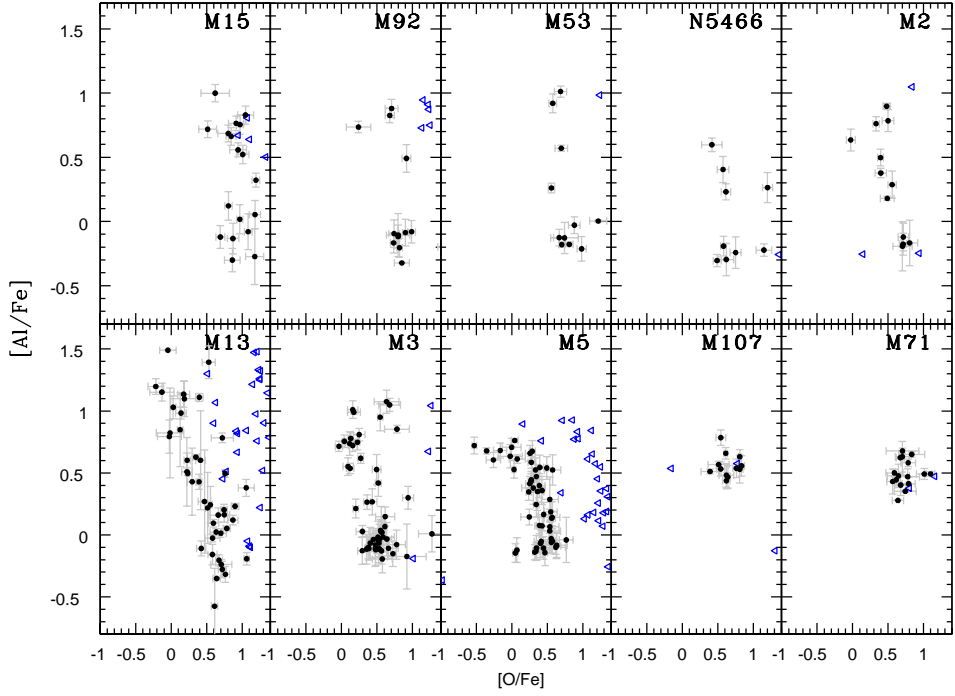}
   \centering
   \caption{Al abundances as a function of O for all GCs in our sample. Black dots are measurements and blue triangles are upper limits in O. The anti-correlation between those elements can be clearly seen for all clusters individually, except M~107 and M~71. }
              \label{fig:AlvsO}%
    \end{figure*}
\section{Discussion}\label{sec:Discussion}
Final abundances, number of lines used, upper limits and random errors are reported in Table~\ref{tab:abund}. 
For some stars, the Fe~I lines were too weak to be detected and only upper limits could be obtained for the metallicity. For those stars, the literature metallicity values from \citet{Harris2010} have been adopted to build the plots. All plots in the following section display random errors.                               
   \begin{figure*}[!ht]
	\includegraphics[angle=-90,width=0.8\textwidth]{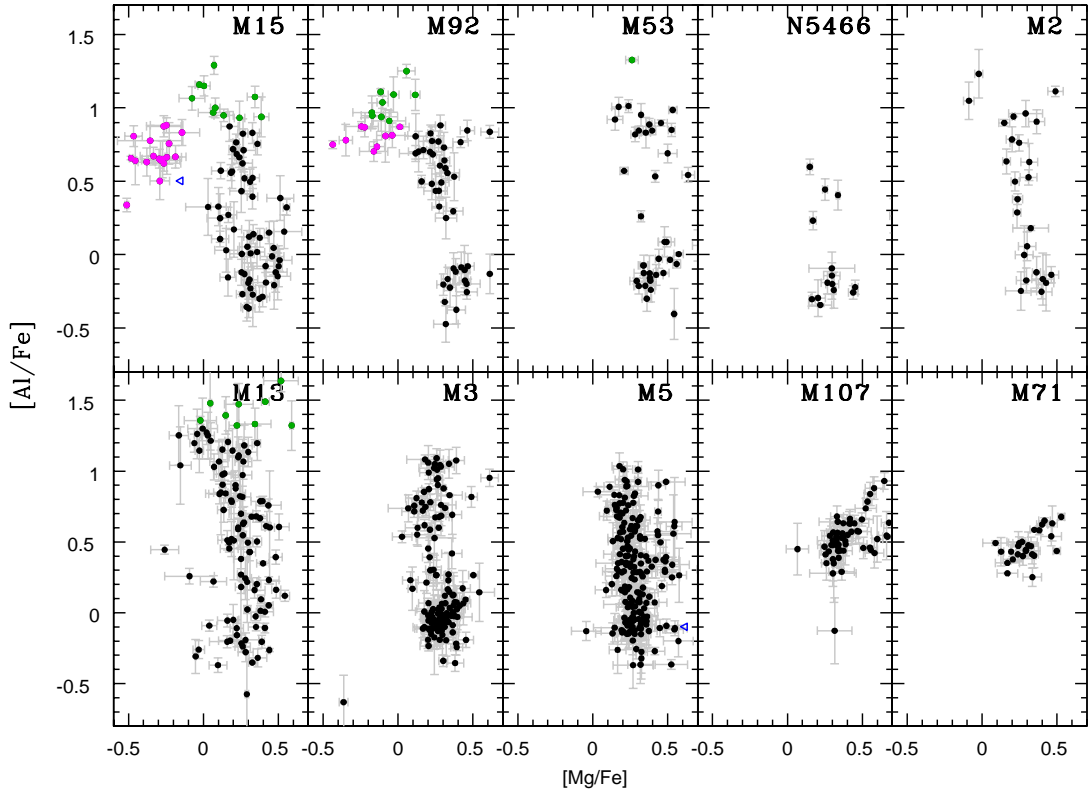}
   \centering
   \caption{Al abundances as a function of Mg abundances for all GCs in our sample. The green points show the Si- and Al-enhanced stars from Fig.~\ref{fig:AlvsSi}. Magenta points highlight the extreme Mg-depleted stars. These stars also show a relatively lower Al abundance.}
              \label{fig:AlvsMg}%
    \end{figure*}

\subsection{Na-O and O-Al anti-correlations}

Na is very difficult to measure in the APOGEE spectra of metal-poor stars because the Na~I lines are very weak. Actually, those lines appear only in the coolest stars of the most metal-rich clusters (hence mostly M~71 and M~107 and very few in M~5, M~3 and M~13). Given that O is also difficult to measure for the warmest stars, our version of the O-Na anti-correlation diagram (Fig.~\ref{fig:NavsO}) appears quite unpopulated. Nevertheless, the values are consistent with very high resolution optical studies \citep{Carretta2009NaO} and thus confirm the high quality of the APOGEE data and of our analysis.

However, the H band contains very clear Al~I lines that are more easily measurable than the Na~I lines. Therefore, Al represents a more robust abundance indicator in the APOGEE data for the study of the multiple stellar population phenomenon in GCs, as first presented by \citet{Meszaros2015}. Indeed, in Fig. \ref{fig:AlvsO} we demonstrate  that we can observe the Al-O anti-correlation for the ten clusters separately, except for M~107 and M~71. Given that those two latter GCs are the most metal-rich ones of the sample, this may indicate that the temperature conditions in the polluters of those two clusters are too low to efficiently produce Al, as illustrated in the case of massive AGB stars polluters by \citet{Ventura2016} and \citet{DellAgli2018}.  
   \begin{figure*}
	\includegraphics[angle=-90,width=0.8\textwidth]{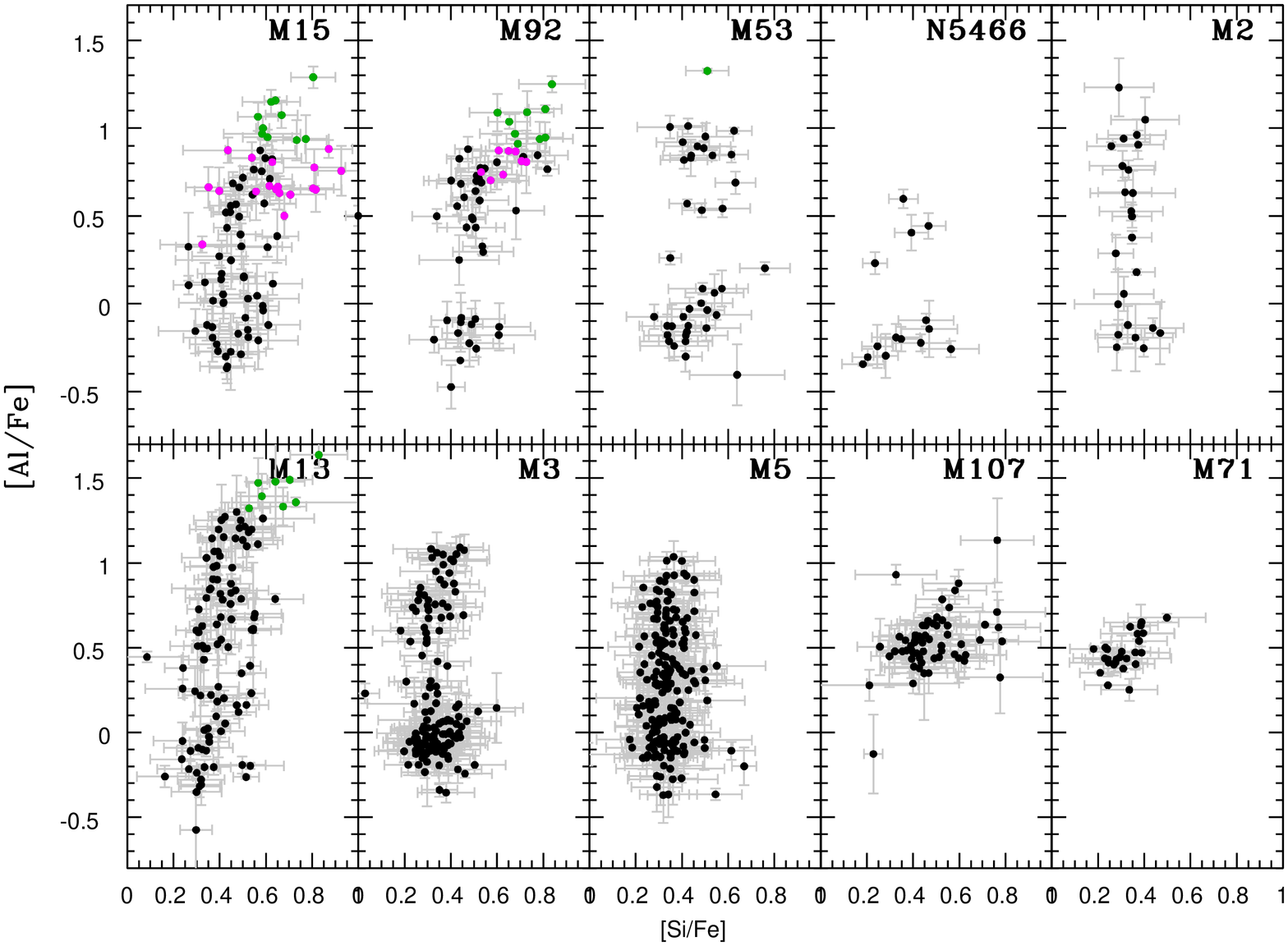}
   \centering
   \caption{Al abundances as a function of Si abundances for all GCs in our sample. The magenta points highlight the very low Mg and Al-weak stars from Fig.~\ref{fig:AlvsMg}. The green points show the Si- and Al-enhanced stars. There is generally no correlation and Si is very homogeneous except in M~15,  M~92 and possibly M~13.}
              \label{fig:AlvsSi}%
    \end{figure*}
   \begin{figure}
	\includegraphics[angle=-90,width=0.95\columnwidth]{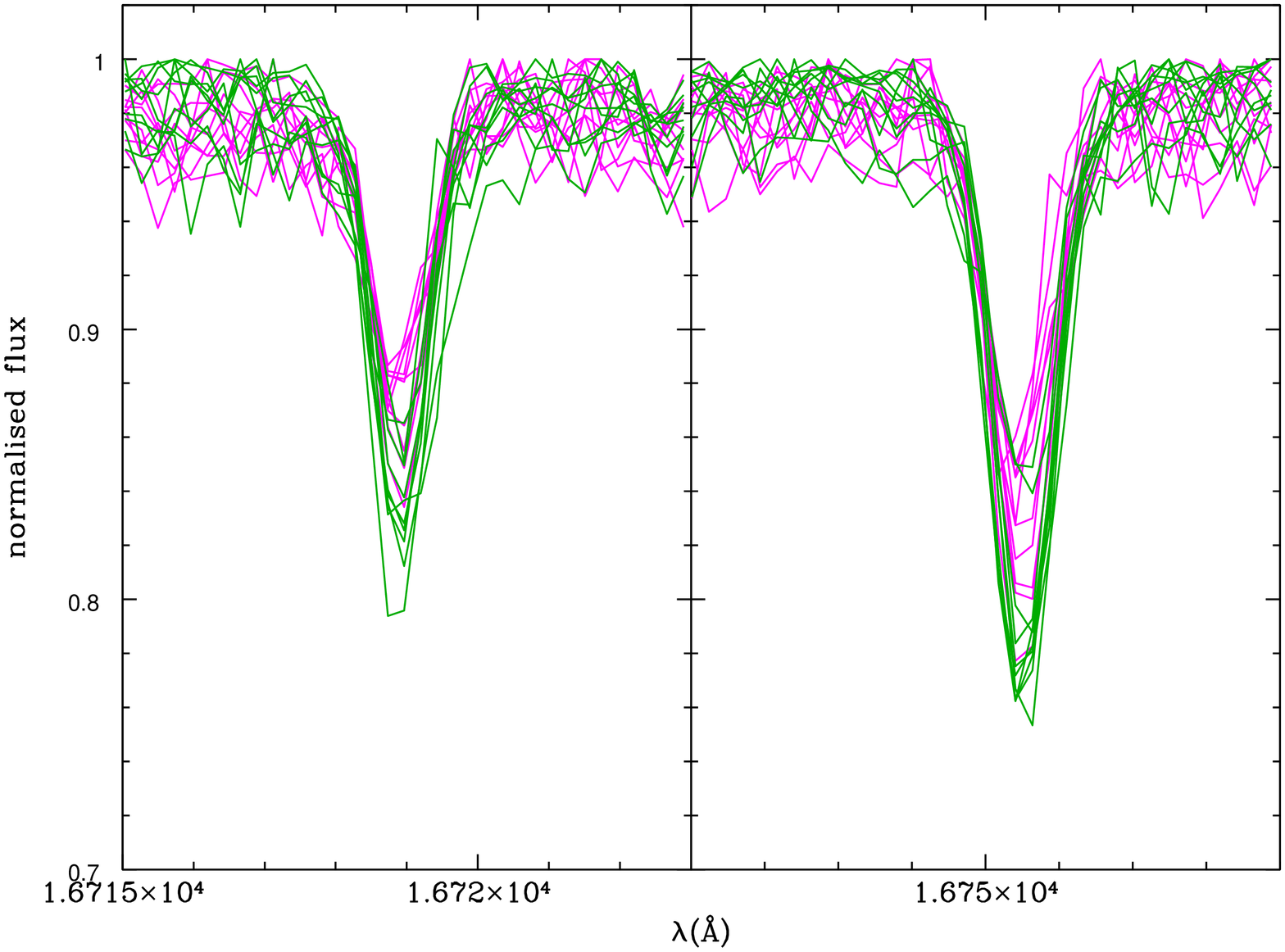}
   \centering
   \caption{Al~I lines in M15 stars.The green lines show the Si-enhanced stars from Fig.~\ref{fig:AlvsSi}, and the magenta lines show the Mg- and Al-weak stars from Fig.~\ref{fig:AlvsMg}.}
              \label{fig:Allines_M15}%
    \end{figure}

\subsection{Mg, Al and Si}\label{sec:MgAlSi}

In Fig.~\ref{fig:AlvsMg}, we establish the Mg-Al anti-correlation for all clusters  as already found in literature \citep[except M~107 and M~71 as already noted by][]{Meszaros2015}. This implies that the Mg-Al chain is active in almost all the clusters (including M5) in contrast to the conclusions of \citet{Carretta200917clusters}. An interesting aspect shown in these data when scrutinizing those diagrams, is that stars with [Mg/Fe] $<$ 0 in M~15 and  M~92 seem to have extremely depleted Mg and have lower Al abundances than one would expect from the extension of the Mg-Al anti-correlation for stars with such low [Mg/Fe] abundances. We stress that, regarding any possible analysis bias, we have not been able to find any dependence on effective temperature nor on evolutionary status (RGB or eAGB/RHB). Moreover, in Fig.~\ref{fig:Allines_M15} we show the spectra around the two Al~I lines present in the APOGEE spectra. This figure unambiguously demonstrates that the Al~I lines are weaker in the most Mg-depleted stars. Thus, we conclude that the relatively low Al in such extremely Mg-depleted stars in M~15 and M~92 is real.

The Al-Si plane can be seen in Fig.~\ref{fig:AlvsSi}. Overall Si is constant and consistent with field star value for similar metallicities. But three clusters (M15,  M~92 and M~13) show a significant Si enhancement for the Al richest stars. \citet{Yong2005}, \citet{Carretta200917clusters} and \citet{Meszaros2015} interpreted the Al-Si correlation as a signature of $\rm ^{28}Si$ leakage from the Mg-Al chain. Interestingly, most of the Si enriched stars in M~15 and  M~92 also seem to correspond to the extreme Mg-depleted and midly enhanced Al stars mentioned above.  According to \citet{Prantzos2017} during H-burning processes, Si begins to be produced above 80MK. According to the same authors $^{27}$Al is expected to be progressively produced up to $\sim$80~MK but begins to be depleted above 80~MK. Therefore, temperatures in polluters above 80~MK would explain satisfactorily the existence of Mg- and Al-weak (and Si enhanced) stars in M~15 and M~92. 

Interestingly, M~13 also shows at least as high Al enhancements as M~15 and  M~92 ([Al/Fe] > 1).  While only a few stars may show some Si enhancement, none of them appears to be extremely Mg-depleted and weakly enhanced in Al such as in M~15 and  M~92. Knowing that M~13 is more metal-rich than M~15 and  M~92, this may suggest that there is a metallicity dependence on the production/depletion yields of the Al source.

Finally, if we now consider the whole sequence formed by the Mg-depleted stars with the more standard Mg-Al anti-correlation, its appearance is similar to a hook, because the most Mg-depleted stars have smaller than expected Al abundances. Although our data is certainly one of the most extensive spectroscopic studies of those clusters, some data may still miss, and the difficulty of analyzing such low metallicity stars can possibly result in biases in the trends. On the other hand, if the most Mg-poor stars do indeed have lower than expected Al abundances, this would imply that standard correlation and anti-correlation patterns observed in GCs are in fact more complex than previously thought. This is well in line with the recent partitioning of the clusters photometric map (the so-called chromosome map) into various sub-populations by  \citet{Milone2017} and in particular the discovery of the puzzling extension of the first population by \citet{Lardo2018}. 

\subsection {C and N}
   \begin{figure*}
	\includegraphics[angle=-90,width=0.8\linewidth]{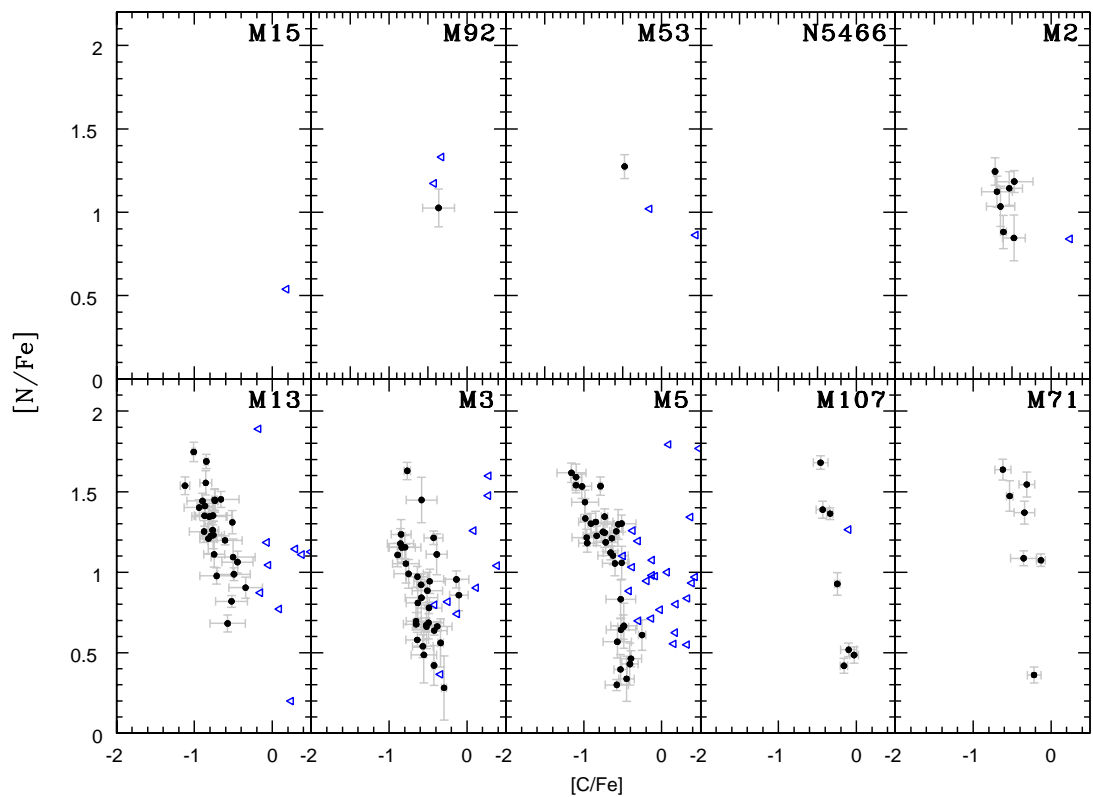}
   \centering
   \caption{N abundances as a function of C abundances for all GCs in our sample. Black dots are measurements and blue triangles are upper limits in C. The N-C anti-correlation is clearly observed for the most metal-rich clusters (bottom panels).}
              \label{fig:NvsC}%
    \end{figure*}
\begin{figure}
	\includegraphics[angle=-90,width=\columnwidth]{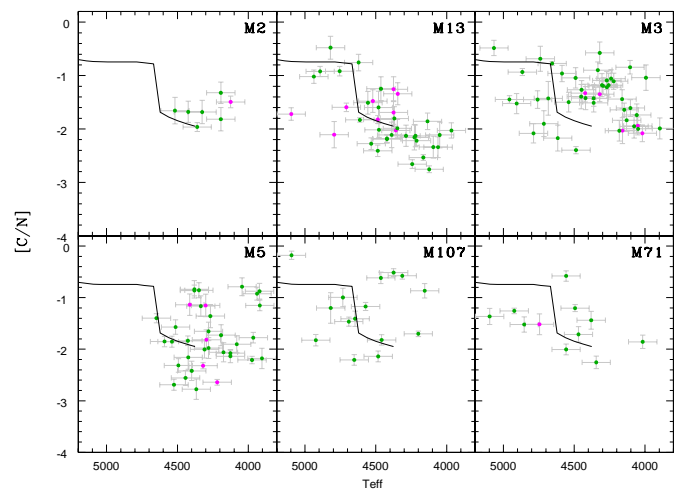}
   \centering
   \caption{[C/N] ratios as a function of effective temperature for the most metal-rich GCs of our sample. RHB/eAGB stars are in magenta, while RGB stars are in green. The data are compared with a \citet{Lagarde2012} 0.85$\rm M_\odot$ Z=0.0001 model (solid black line). Note that this model has been shifted by -0.5 in C/N to match approximately the post-first dredge up C/N data value. }
              \label{fig:CNvsTeff}%
    \end{figure}

In Fig.~\ref{fig:NvsC}, the C and N anti-correlation can be observed for most of the clusters except the most metal-poor ones where CO and CN lines are too weak to be detected. 
We first wondered about the influence of the intrinsic red giant branch extra-mixing over the C and N data. Indeed, the existence of the extra-mixing and its C and N signature is clearly seen in the APOGEE field stars \citep{Masseron2017,Shetrone2018} and can have a very large impact on the C and N yields at very low metallicity. We plot in Fig. \ref{fig:CNvsTeff} the C/N ratios over the effective temperature and compare to the model expectations using a prescription for the extra-mixing from \citet{Lagarde2012}. The model predicts a significant drop of the C/N ratio after the bump luminosity around 4700~K. However, it is difficult to evaluate the impact of the extra-mixing on the C and N yields by comparing the C and N abundances before and after the  luminosity bump because the [C/N] data show a very large scatter. Indeed, it is very likely that the extra-mixing signature depends also on the initial abundances of C and N, which have been proven to vary greatly in globular clusters from the observation of unevolved stars \citep[see][and references therein]{Gratton2004}. Consequently, we stress that only field stars represent reliable test-beds for studying extra-mixing along the RGB but no GCs stars should be used to do so, in contrast to \citet{Lagarde2018}'s work. 

However, extra-mixing is believed to induce CN-cycling, i.e. that extra-mixing is affecting C and N yields but not affecting the C+N+O yields. By looking at the C+N+O yields we can verify if the C+N+O is varying in a cluster, and thus constrain the nucleosynthesis of the cluster polluters independently of extra-mixing effect. We have checked the C+N+O values in our clusters. Compared to \citet{Meszaros2015}, it appears quite clear now that no correlation exists between C+N+O and Al for at least three clusters (M~13, M~3 and M~5). Furthermore, the C+N+O data are consistent with no variations within the errors and are also consistent with field stars values of similar metallicities \citep[e.g. ][]{Gratton2000}. Therefore, the CNO cycle is also occurring in the GCs polluters. Moreover, we do not find significant enhancement in any of our clusters (including those with upper limits) compared to the value of field stars such as it has been claimed by \citet{Yong2009,Yong2015} in NGC~1851, although disputed by \citet{Villanova2010}.   

\subsection{Correlations with cluster global properties}\label{sec:spread}
   \begin{figure}[h!]
	\includegraphics[angle=-90,width=0.95\columnwidth]{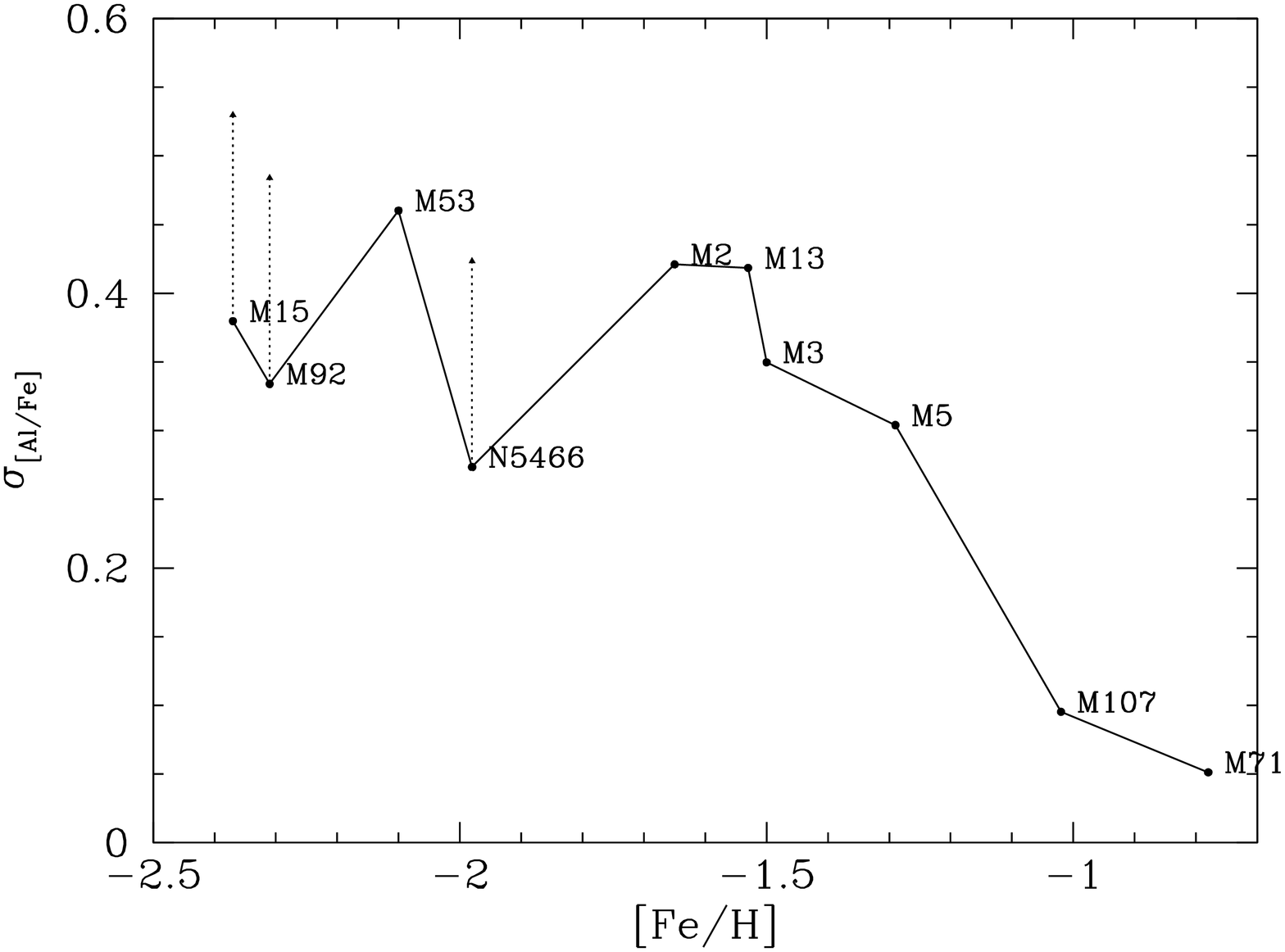}
	\includegraphics[angle=-90,width=0.95\columnwidth]{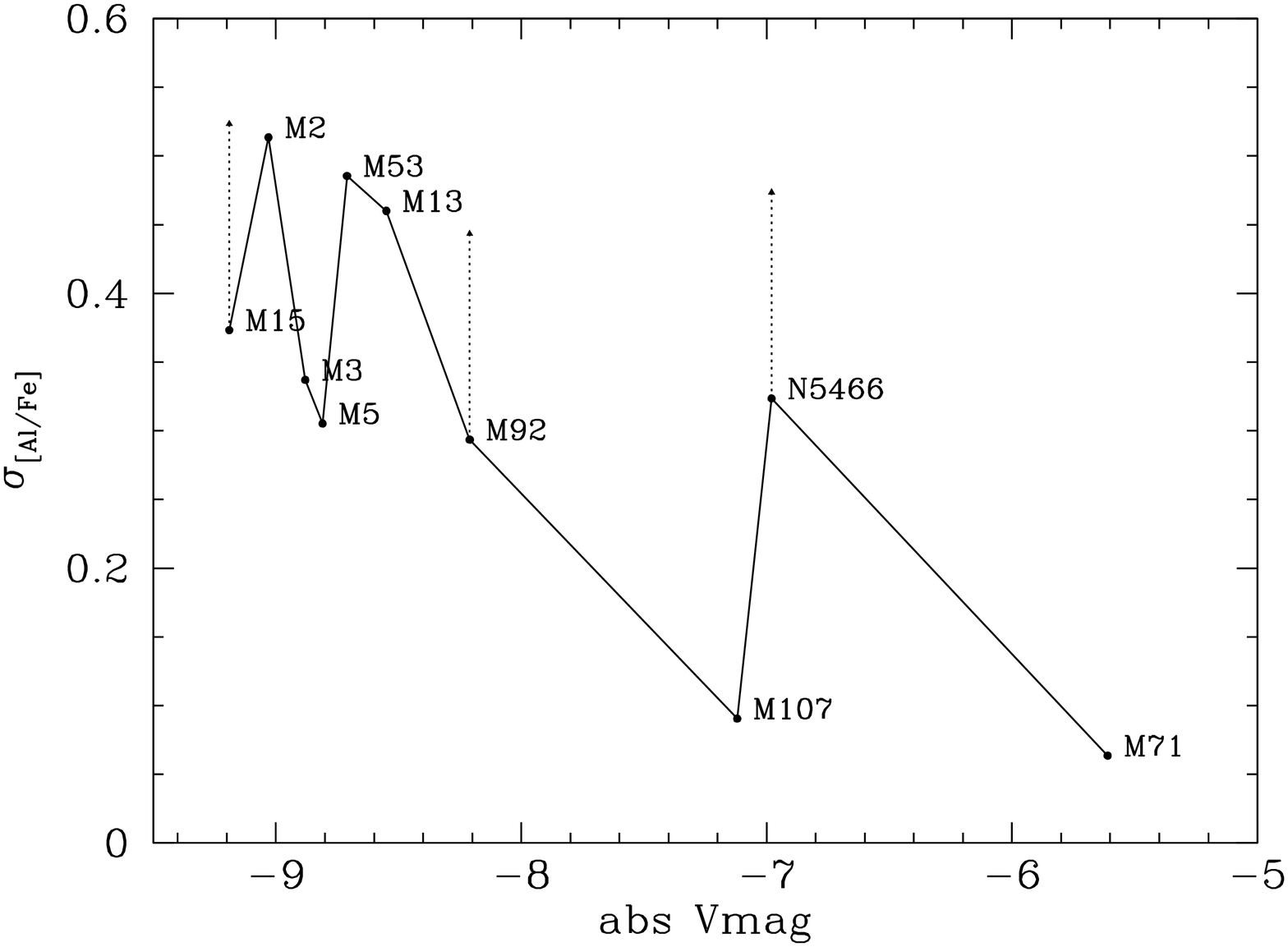}
	\includegraphics[angle=-90,width=0.95\columnwidth]{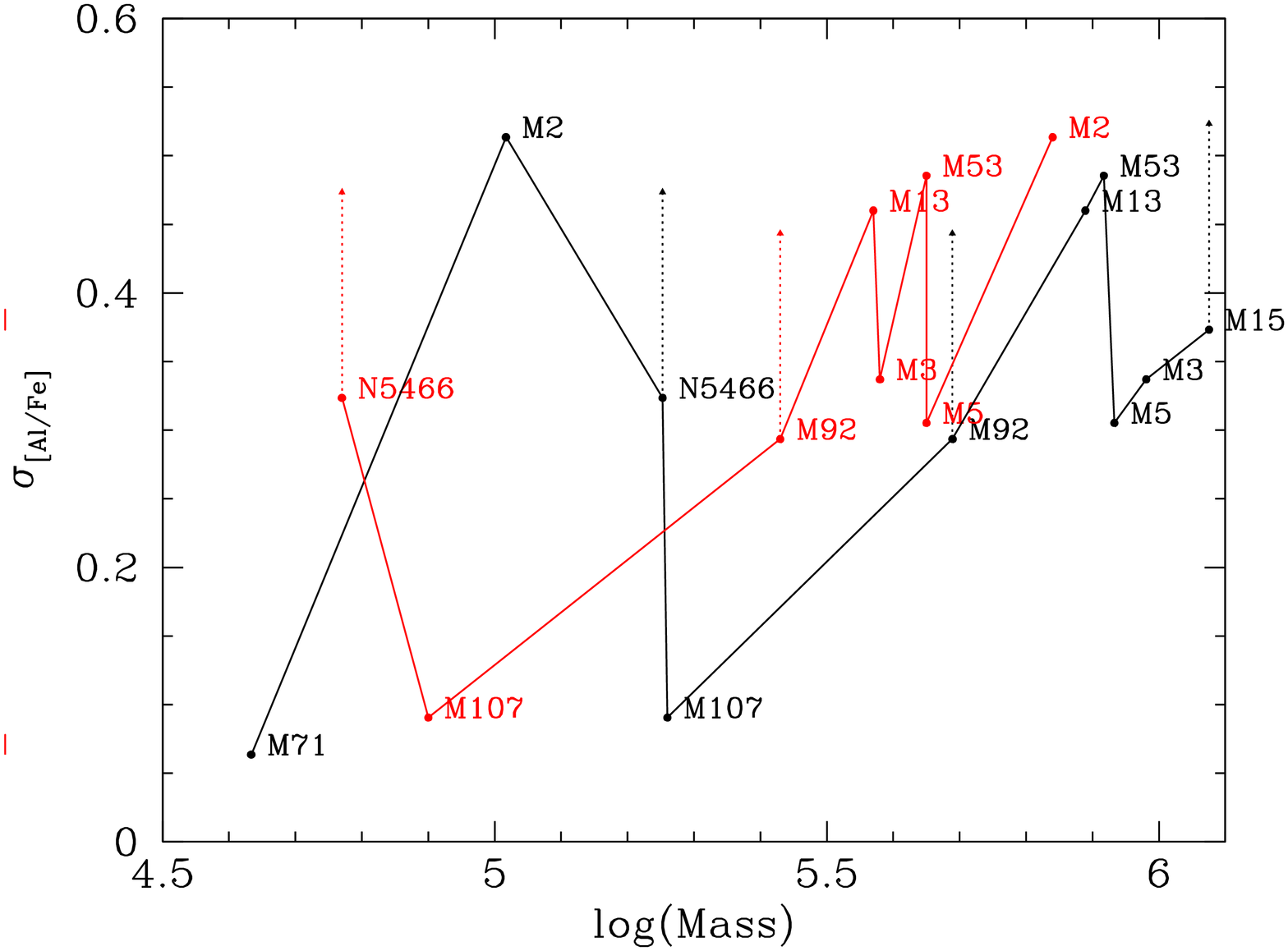}
   \centering
   \caption{The spread in Al as a function of global cluster parameters (metallicity, absolute magnitude and total masses). M~15,  M~92 and NGC~5466 have probably underestimated spread and have been highlighted accordingly. The metallicity and the absolute magnitudes are extracted from \citet{Harris2010}. The total masses are computed by \citet{McLaughlin2005} (black line) and from a compilation by \citet{Boyles2011} (red line). There is probably an anti-correlation with metallicty and maybe a correlation with absolute magnitude.}
              \label{fig:Alspread}%
    \end{figure}

   \begin{figure*}
	\includegraphics[angle=-90,width=0.8\linewidth]{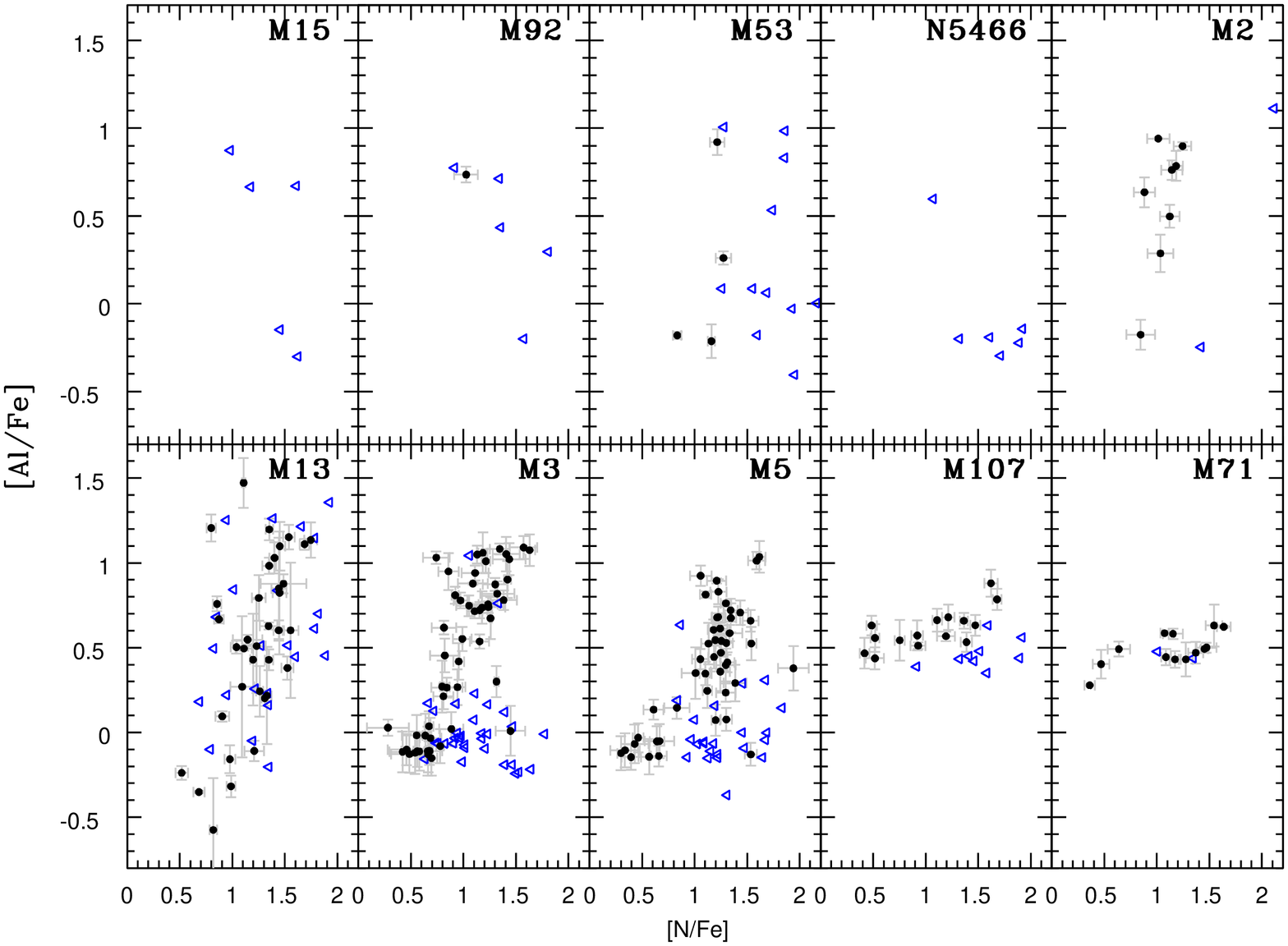}
   \centering
   \caption{Al abundances as a function of N abundances for all the clusters sorted by metallicity. Blue triangles are upper limits in N. Al and N are correlated, but the Al spread is decreasing with increasing metallicity.}
              \label{fig:AlvsN}%
    \end{figure*}

In the previous sections, we have confirmed that Al is (anti-)~correlated with many elements (C, N, O, Na, Mg and Si). Given the completeness of the Al measurements in our sample as well as its large variations, we consider this element as the best representation to evaluate the extent of the cluster's multiple populations.
As already suggested by \citet{Carretta200917clusters} \citep[but see also][]{Meszaros2015,Ventura2016,DellAgli2018}, Al spread within each cluster decreases with increasing cluster metallicity. Thanks to our large sample, we can now attempt to compare quantitatively this spread against clusters metallicities as well as other global properties. In Fig.\ref{fig:Alspread} , we plot the Al spread (derived from the deviation from the median) against cluster metallicity, absolute magnitude and mass. We remind that M~15 and  M~92 probably have lower than expected Al abundance in the most extreme cases. Therefore, the Al spread may be underestimated for those clusters. We also remark that NGC~5466 has a quite low number of stars, thus we also probably under-evaluate the real Al spread. The diagrams presented Fig.~\ref{fig:Alspread} suggest an anti-correlation between the Al spread and the metallicity, as well as possible correlation between the Al spread and absolute magnitude.  From a nucleosynthetic point of view, this potentially implies that the Mg-Al chain reaction is becoming less important with increasing metallicity and/or cluster luminosity. Moreover, although cluster absolute luminosity is known to be a proxy for cluster mass, we found that the correlation of the Al spread with cluster mass is quite uncertain, mostly because cluster mass determination is model dependent.\\
Similarly, \citet{Carretta2009NaO} found a bilinear anti-correlation between their Na$\rm_{max}$ ($\propto$ Na spread) and clusters luminosities and metallicities. This is consistent with our results because Al and Na have been demonstrated to be correlated \citep{Carretta200917clusters}. But this is contrasting with \citet{Milone2017} who rather found a correlation between metallicity - as well as cluster magnitude - and the width of the RGB. This apparent contradiction could be explained by the fact that the RGB width in a colour-magnitude diagram is known to be sensitive to N because of molecular bands (but not to Al which has no molecular bands). Thus, we can deduce that in the \citet{Milone2017}'s work, it is the N spread which is increasing with metallicity. But this phenomenon is difficult to be confirmed by our data ( Fig.~\ref{fig:AlvsN}) because N is measured in only a fraction of the stars of our sample, statistically weakening the possibility of measuring accurately the N spread. Actually, in all the clusters we could measure N in a large enough number of stars (M~13, M~3, M~5, M~107 and M~71) the N spread seems rather constant ($\sim$ 1.5dex) except for the most metal-poor one M~2 where it is nearly null.

\subsection{K}
   \begin{figure}[!h]
	\includegraphics[angle=-90,width=\linewidth]{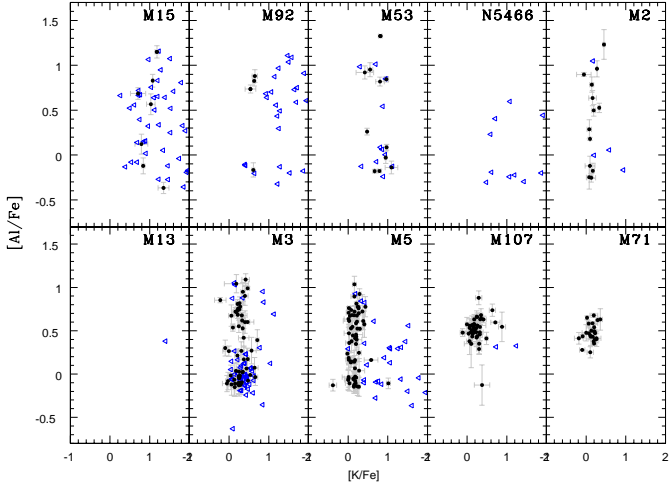}
   \centering
   \caption{Al abundances as a function of K abundances for all GCs in our sample. Blue triangles are upper limits in K.  While K is difficult to measure in very low metallicity GCs, there is no significant spread in the remaining ones. Note that M~13 does not contain measurements because the K lines at the cluster doppler shift fall in one of the APOGEE CCD gap where no measurement can be made.}
              \label{fig:AlvsK}%
    \end{figure}

Although the exact temperature for the onset of the K production by H-burning nucleosynthesis may be still debated \citep[120-180~MK][]{Ventura2012,Iliadis2016}, the temperature must be much higher than the one for the onset of Al or Si production($\sim$80~MK). We certainly observe Si production in M~15 and  M~92 (Fig.~\ref{fig:AlvsSi}), indicating that in these clusters the polluters had reached the highest temperatures. But neither in those extreme clusters -nor in the others- is a variation in K observed (Fig.\ref{fig:AlvsK}). We conclude that the K measured in those clusters is consistent with no K production, thus relatively low-temperature nucleosynthesis conditions (but still high enough to produce Si), and that K production in GCs is rare, in agreement with the conclusions of  \citet{Takeda2009} and \citet{Carretta2013}. Still, it has been clearly established by \citet{CohenKirby2012} and \citet{Mucciarelli2015} that NGC~2808 and NGC~2419 show K enhancement correlated with notably Al. It is remarkable that these two clusters have metallicities in the same range as our sample. Therefore, in contrast to Si production (Sec.\ref{sec:MgAlSi}) or Al spread (Sec.\ref{sec:spread}), K production in GCs is not only a function of metallicity. In any case, the absolute luminosities of those K-enhanced clusters are among the largest. This may corroborate the idea that the clusters peculiar chemistry is not only a function of metallicity but also a function of cluster luminosity \citep{Carretta200917clusters,Milone2017}

\subsection{Ca}
   \begin{figure}[!h]
	\includegraphics[angle=-90,width=\linewidth]{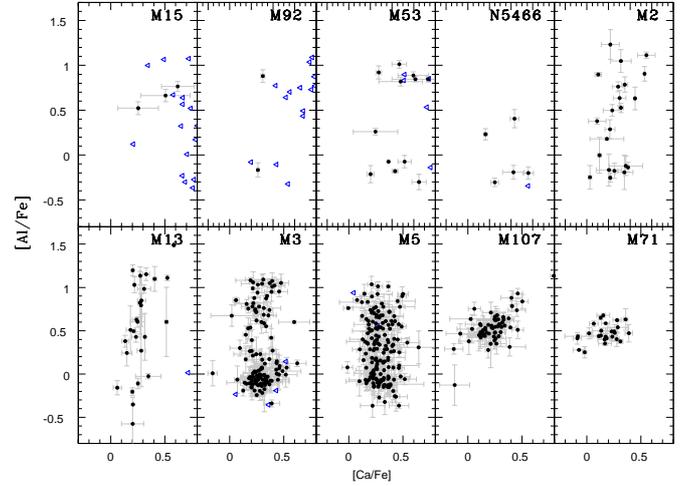}
   \centering
   \caption{Al abundances as a function of Ca abundances for all GCs in our sample. Blue triangles are upper limits in Ca. While Ca is difficult to measure in very low metallicity GCs, there is no significant spread in any of the clusters.}
              \label{fig:AlvsCa}%
    \end{figure}
\citet{Marino2009} observed a Ca spread in M~22. But Ca is not expected to be affected by H-burning processes \citep{Prantzos2017}. Fig.~\ref{fig:AlvsCa} shows the Ca abundances measured in our 10 GCs. It is clear in that figure that no correlation with Al is observed. The star-to-star scatter is also small enough to show no enhancement in any of the clusters. Therefore, M~22 remains certainly an exceptional cluster regarding Ca enhancement, although \citet{Mucciarelli2015NLTE} suspect that NLTE effects could be also responsible for such an observational spread.

\subsection{s-process elements: Ce and Nd}
   \begin{figure}
	\includegraphics[angle=-90,width=\columnwidth]{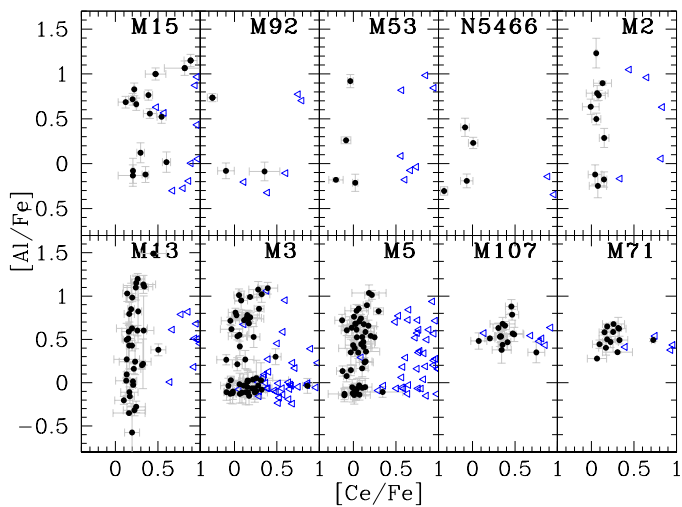}
   \centering
   \caption{Al abundances as a function of Ce abundances for all GCs in our sample. Black dots are measurements and blue triangles are upper limits in Ce. Except for M~15 (and perhaps  M~92), all clusters are consistent with a homogeneous Ce abundance.}
              \label{fig:AlvsCe}%
    \end{figure}
   \begin{figure}
	\includegraphics[angle=-90,width=\columnwidth]{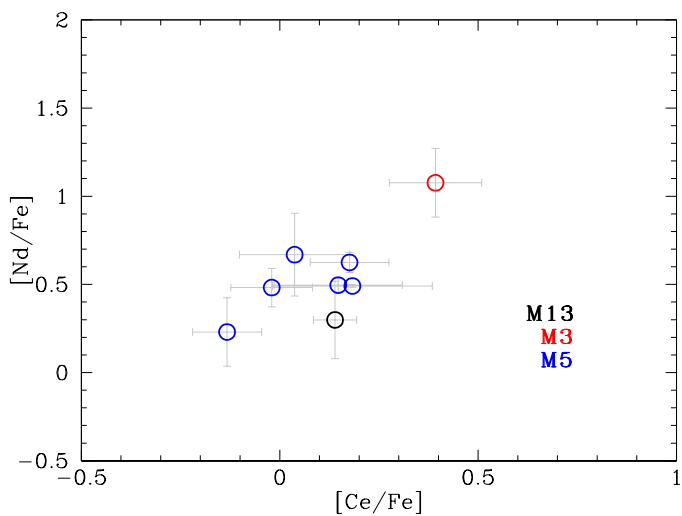}
   \centering
   \caption{Nd abundances as a function of Ce abundances for all GCs in our sample. Although there are few measurements of Nd, they correlate well with Ce.}
              \label{fig:NdvsCe}
    \end{figure}

In Fig.~\ref{fig:AlvsCe}, we confirm that most GCs do not seem to show a significant spread in Ce nor show any enhancement compared to field stars value as already remarked by \citet{Gratton2004}. In contrast, M~15 shows a significant spread. Nd is difficult to measure and we obtain some constraining values only for the most favorable spectra, i.e. in the coolest giants of the most metal-rich GCs. We still observe a consistent correlation with Ce (Fig.~\ref{fig:NdvsCe}). Unfortunately, with  only one neutron-capture element measured in M~15, it is not possible to assign the nucleosynthetic origins in those stars, and notably disentangle a r-process and a s-process  nucleosynthesis such as discussed in  M~22 and  M~2 \citep{Roederer2011,Yong2014}. Therefore, our conclusions concerning M~15 are bound to \citet{Sobeck2011}, i.e. that the Ce we have measured has probably a pure r-process origin. We also agree that the Ce dispersion in M~15 is comparable to that of the halo field at the same metallicity, although we have now clearly established that M~15 also shows large Al variation. But we hardly see a correlation with Al, confirming that polluters in our GCs do not produce Ce and is probably from the GCs primordial formation gas. 

\section{Conclusions}
We confirm that APOGEE spectra provide precise elemental abundances for many elements. However, we show that the latest data release (DR14) still suffers from uncertainties, in particular regarding extreme abundances such as those found in some cases for C and N, as already pointed out by \citet{Jonsson2018}, but more generally for very low metallicity spectra ([Fe/H]$<$-1.5) in which metallic lines become very weak and are extremely sensitive to choice of prescription for parameters like macroturbulence and NLTE effects. We emphasize the crucial lack of an upper limits flagging system fro APOGEE. Finally, we demonstrate that the ASPCAP DR14 T$\rm_{eff}$ are probably biased whenever [O/$\alpha$] are not solar.

With our independent analysis, we measure almost all the elemental abundances needed for GCs studies, (newly-included the neutron-capture elements Ce and Nd in this survey), except He and Na. Although those latter two elements abundances would certainly be very interesting to be measured as well, we demonstrate that all known light elements anti-/correlation can be recovered with a high precision. Consequently, we confirm that H-burning reactions are the main nucleosynthesis processes that have occurred in all the clusters in this analysis.  

Moreover, along with literature, we have collected some corroborating evidences suggesting that cluster luminosity and metallicity are the two main parameters which are driving the various GCs chemical patterns.  Unfortunately, our sample of GCs is such that we can hardly disentangle what is the main factor controlling the amplitude of the pollution between metallicity and cluster magnitude. Indeed, in our sample both are correlated, so that the most metal-rich clusters are also the less luminous, and conversely, the most metal-poor clusters are, on average, the brightest ones. To validate those main dependencies, it will be interesting to analyze more peculiar clusters such as NGC~2808, NGC~2419, M~2, M~4 M~22, NGC~1851, $\omega$~Cen or any young massive stellar clusters planned to be observed by the ongoing SDSS~IV/APOGEE-2 survey.

 Furthermore, we discovered some puzzling stars extremely depleted in Mg and weakly enhanced in Al  that seem to occur only in our most metal-poor clusters, suggesting that the temperature conditions reached in the corresponding polluters are high enough to start burning Al.  Finally, in the same Mg-Al plane, we have observed that the data are forming a hook.  Any model that is attempting to explain the multiple populations in GCs must now be able to self-consistently account for such a turnover in the Mg-Al anti-correlation as seen in these new detailed and extensive observations.

\begin{acknowledgements}
We are very grateful for the fruitful comments from P. Ventura and the help from S. Villanova.

T.M. acknowledges support from Spanish  Ministry  of  Economy and Competitiveness (MINECO) under the 2015 Severo Ochoa Program SEV-2015-0548. T.M., D.A.G.H., O.Z., and F.D.A. also acknowledge support by  the MINECO under grant AYA-2017-88254-P.
  SzM has been supported by the Premium Postdoctoral Research Program of the Hungarian Academy of Sciences, and by the Hungarian NKFI Grants K-119517 of the Hungarian National Research, Development and Innovation Office. H. J. acknowledges support from the Crafoord Foundation, Stiftelsen Olle Engkvist Byggm\"astare, and Ruth och Nils-Erik Stenb\"acks stiftelse. D.G. acknowledges support from the Chilean Centro de Excelencia en Astrof\'isica y Tecnolog\'ias Afines (CATA) BASAL grant AFB-170002.
D.G. also acknowledges financial support from the Dirección de Investigación y Desarrollo de
la Universidad de La Serena through the Programa de Incentivo a la Investigación de
Académicos (PIA-DIDULS). T.C.B. acknowledges partial support from grant PHY 14-30152; Physics Frontier Center/JINA Center for the Evolution of the Elements (JINA-CEE), awarded by the US National Science Foundation.

This paper made use of the IAC Supercomputing facility HTCondor (http://research.cs.wisc.edu/htcondor/), partly financed by the Ministry of Economy and Competitiveness with FEDER funds, code IACA13-3E-2493.

Funding for the Sloan Digital Sky Survey IV has been provided by the Alfred P. Sloan Foundation, the U.S. Department of Energy Office of Science, and the Participating Institutions. SDSS acknowledges support and resources from the Center for High-Performance Computing at the University of Utah. The SDSS web site is www.sdss.org.

SDSS is managed by the Astrophysical Research Consortium for the Participating Institutions of the SDSS Collaboration including the Brazilian Participation Group, the Carnegie Institution for Science, Carnegie Mellon University, the Chilean Participation Group, the French Participation Group, Harvard-Smithsonian Center for Astrophysics, Instituto de Astrofísica de Canarias, The Johns Hopkins University, Kavli Institute for the Physics and Mathematics of the Universe (IPMU) / University of Tokyo, the Korean Participation Group, Lawrence Berkeley National Laboratory, Leibniz Institut für Astrophysik Potsdam (AIP), Max-Planck-Institut für Astronomie (MPIA Heidelberg), Max-Planck-Institut für Astrophysik (MPA Garching), Max-Planck-Institut für Extraterrestrische Physik (MPE), National Astronomical Observatories of China, New Mexico State University, New York University, University of Notre Dame, Observatório Nacional / MCTI, The Ohio State University, Pennsylvania State University, Shanghai Astronomical Observatory, United Kingdom Participation Group, Universidad Nacional Autónoma de México, University of Arizona, University of colourado Boulder, University of Oxford, University of Portsmouth, University of Utah, University of Virginia, University of Washington, University of Wisconsin, Vanderbilt University, and Yale University.

\end{acknowledgements}

\bibliographystyle{aa}
\bibliography{NdCeClusters} 

\begin{table*}
\begin{tabular}{llllllllllllll}
\hline \\ 
star                &   cluster & status\tablefootmark{*} & Teff    &    logg  &     [Fe/H] & $\sigma$ &  [C/Fe] & $\sigma$ & \# &  [N/Fe] & $\sigma$ & \# & $\cdots$\\
\hline \\ 
2M15184612+0204467 &       M5 &       HB &     7054 &     3.01 & $\cdots$ & $\cdots$ & $\cdots$ & $\cdots$ & 0 & $\cdots$ & $\cdots$ & 0 & $\cdots$  \\ 
2M15184730+0207253 &       M5 &     eAGB &     4998 &     1.70 &   -1.314 &    0.129 &    0.182 &    0.171 & 1 &   <1.279 & $\cdots$ & 1 & $\cdots$  \\ 
2M15185167+0201178 &       M5 &     eAGB &     5061 &     1.81 &   -1.358 &    0.046 & $\cdots$ & $\cdots$ & 0 & $\cdots$ & $\cdots$ & 0 & $\cdots$  \\ 
2M15185197+0211217 &       M5 &       HB &     5681 &     2.02 &   -1.575 &    0.345 &    1.554 &    0.094 & 1 &   <2.578 & $\cdots$ & 1 & $\cdots$  \\ 
2M15185499+0205525 &       M5 &       HB &     6366 &     2.30 & $\cdots$ & $\cdots$ & $\cdots$ & $\cdots$ & 0 & $\cdots$ & $\cdots$ & 0 & $\cdots$  \\ 
2M15185515+0214337 &       M5 &     eAGB &     4663 &     1.53 &   -1.236 &    0.048 &   -0.313 &    0.125 & 1 &   <1.131 & $\cdots$ & 1 & $\cdots$  \\ 
2M15185731+0203077 &       M5 &     eAGB &     4995 &     1.90 &   -1.206 &    0.007 &   <1.164 & $\cdots$ & 1 &   <2.519 & $\cdots$ & 1 & $\cdots$  \\ 
2M15190831+0201421 &       M5 &       HB &     5394 &     2.04 &   -1.206 &    0.121 & $\cdots$ & $\cdots$ & 0 & $\cdots$ & $\cdots$ & 0 & $\cdots$  \\ 
2M15191829+0209175 &       M5 &       HB &     5934 &     2.48 & $\cdots$ & $\cdots$ & $\cdots$ & $\cdots$ & 0 & $\cdots$ & $\cdots$ & 0 &  $\cdots$ \\ 
2M15193344+0205072 &       M5 &       HB &     7092 &     3.37 & $\cdots$ & $\cdots$ & $\cdots$ & $\cdots$ & 0 & $\cdots$ & $\cdots$ & 0 &  $\cdots$ \\ 
2M16315197-1303389 &     M107 &      RGB &     4950 &     2.66 &   -1.071 &    0.065 &   -0.326 &    0.279 & 2 &    1.127 &    0.056 & 16 &  $\cdots$ \\ 
2M16315358-1302179 &     M107 &      RGB &     5080 &     2.93 &   -1.069 &    0.133 &    0.009 &    0.286 & 2 &   <0.666 & $\cdots$ & 1 &  $\cdots$ \\ 
2M16315926-1303045 &     M107 &      RGB &     5031 &     2.91 &   -0.999 &    0.063 &   -0.081 &    0.102 & 1 &   <1.726 & $\cdots$ & 1 &  $\cdots$ \\ 
2M16320382-1308186 &     M107 &      RGB &     5009 &     2.93 &   -1.065 &    0.086 &    0.076 &     0.33 & 3 &   <1.445 & $\cdots$ & 1 &  $\cdots$ \\ 
2M16320891-1300161 &     M107 &      RGB &     4773 &     2.19 &   -0.941 &    0.074 &    -0.03 &    0.124 & 5 &   <0.648 & $\cdots$ & 1 &  $\cdots$ \\ 
2M16320904-1302270 &     M107 &      RGB &     5003 &     2.66 &   -0.953 &    0.026 &   -0.322 &    0.196 & 2 &    1.435 &    0.051 & 14 &  $\cdots$ \\ 
2M16321394-1301086 &     M107 &      RGB &     4307 &     1.55 &   -1.044 &    0.021 &   -0.172 &    0.027 & 4 &    0.354 &    0.058 & 13 &  $\cdots$ \\ 
2M16322086-1302131 &     M107 &      RGB &     4937 &     2.65 &   -0.826 &    0.062 &  <-0.057 & $\cdots$ & 1 &   <1.728 & $\cdots$ & 1 &  $\cdots$ \\
$\cdots$ & $\cdots$ & $\cdots$ & $\cdots$ & $\cdots$ & $\cdots$ & $\cdots$ & $\cdots$ & $\cdots$  & $\cdots$ & $\cdots$  & $\cdots$   & $\cdots$   & $\cdots$  \\
\hline \\ 
\end{tabular}
\tablefoot{
\tablefootmark{*}{eAGB are distinguished from HB stars such that T$\rm_{eff}<$5300~K   }
}
\caption{Parameters and abundances for the sample stars. {\it The full table will be available electronically}.} 
\label{tab:abund}
\end{table*}

\end{document}